\documentclass[twocolumn,trackchanges]{aastex631}

\shorttitle{Extended Fe K$\alpha$ Emission in Nearby AGN}
\shortauthors{Masterson \& Reynolds}

\begin{document}

\title{Probing the Extent of Fe K$\alpha$ Emission in Nearby Active Galactic Nuclei using Multi-Order Analysis of Chandra High Energy Transmission Grating Data}

\author[0000-0003-4127-0739]{Megan Masterson}
\affiliation{MIT Kavli Institute for Astrophysics and Space Research,
Massachusetts Institute of Technology, 
Cambridge, MA 02139, USA}
\affiliation{Institute of Astronomy,
University of Cambridge,
Madingley Road,
Cambridge CB3 0HA, UK}
	
\author[0000-0002-1510-4860]{Christopher S. Reynolds}
\affiliation{Institute of Astronomy,
University of Cambridge,
Madingley Road,
Cambridge CB3 0HA, UK}

\correspondingauthor{Megan Masterson}
\email{mmasters@mit.edu}

\begin{abstract}

We present a study of the narrow Fe K$\alpha$ line in seven bright, nearby AGN that have been observed extensively with the Chandra High Energy Transmission Grating (HETG). The HETG data reveal a wider Fe K$\alpha$ line in the first order spectrum than in the second and third order spectra, which we interpret as the result of spatially extended Fe K$\alpha$ emission. We utilize these differences in narrow Fe K$\alpha$ line widths in the multi-order Chandra HETG spectra to determine the spatial extent and intrinsic velocity width of the emitting material in each object. We find that there is modest evidence for spatially extended emission in each object, corresponding to extension of $r\sim5-100$ pc. These distances are significantly larger than those inferred from velocity widths assuming gravitational motions, which give $r\sim0.01-1$ pc. This implies that either the gas is emitting at a range of radii, with smaller radii dominating the velocity width and larger radii dominating the spatial extent, or that the gas is exhibiting non-gravitational motions, which we suggest would be outflows due to slight excess redshift in the line and velocities that exceed the freefall velocity. We also use the spatial extent information to estimate the mass of the emitting gas by counting fluorescing iron atoms, finding masses on the order of $M_\mathrm{gas}\sim10^5-10^8\,M_\odot$. Future work with observatories like XRISM will be able to extend this study to a larger number of AGN and decrease uncertainties that arise due to the low signal-to-noise of the higher order HETG data.

\end{abstract}



\section{Introduction} \label{sec:intro}

Active galactic nuclei (AGN) are powered by accretion onto a central supermassive black hole (SMBH) with $M_\mathrm{BH} \sim 10^6 - 10^{9.5} M_\odot$. Canonically, the key components of the environment around the SMBH are an accretion disk, the broad-line region (BLR) in which fast moving gas clouds produce broad optical emission lines, a parsec-scale torus responsible for large amounts of obscuration and reprocessing of emission from the central engine \citep[see][for a review]{Hickox2018}, and the narrow-line region (NLR), which extends from beyond the torus into the host galaxy and produces narrow optical emission lines. The torus is an important factor in the unified model of AGN \citep[e.g.][]{Antonucci1993,Urry1995,Netzer2015}, but it is still not well understood how this structure connects with the gas reservoir in the host galaxy through feedback and fueling of the central SMBH.

The X-ray spectrum of an AGN is dominated by two components, a power law continuum with a high energy cut-off, which is a result of UV photons from the accretion disk being Compton up-scattered by the hot corona \citep[T $\sim 10^8-10^9$ K; e.g.][]{Fabian2015,Kara2017,Ricci2017,Tortosa2018,Ursini2019}, and a reflection spectrum, which arises due to reprocessing of coronal emission in the surrounding circumnuclear material \citep{Haardt1991,Haardt1993}. The reflection spectrum is a powerful probe of the geometry, kinematics, and ionization state of the surrounding accretion flow and circumnuclear material \citep{George1991}.

One of the most prominent features in the X-ray reflection spectrum is the Fe K$\alpha$ emission line, as iron has a high cosmic abundance and fluorescent yield. The shape of the neutral iron line probes the location and dynamics of the emitting material. Reflection from the inner accretion disk produces a broad iron line (FWHM $\gtrsim 50,000$ km s$^{-1}$), which is asymmetric due to relativistic effects such as gravitational redshift and the transverse Doppler effect \citep{Fabian1989}. The redward extent of this line can be modeled to determine the innermost stable circular orbit (ISCO) and can therefore be used to probe black hole spin \citep[e.g.][]{Tanaka1995,Brenneman2006,Risaliti2013}. To date, modeling the broad iron line is our best way of measuring black hole spins in AGN \citep[see][for a recent review]{Reynolds2021a}. Reflection from more distant material produces the narrow core of the neutral iron line (FWHM $\lesssim 10,000$ km s$^{-1}$), which is a ubiquitous feature of AGN X-ray spectra with sufficient signal-to-noise \citep[e.g.][]{Kaspi2001,Yaqoob2001,Zhou2005,Fukazawa2011,Hitomi2018}. The exact origin of the narrow Fe K$\alpha$ line is still uncertain, with possible origins including the outer accretion disk, BLR, or torus. Various studies utilizing high resolution X-ray spectroscopy of nearby AGN have revealed a range of Fe K$\alpha$ line widths and no correlation with the optical BLR \citep{Yaqoob2004,Nandra2006,Shu2010}, whereas others utilizing both spectroscopy and variability have found that most of the Fe K$\alpha$ emission is produced within the dust sublimation radius \citep{Gandhi2015,Minezaki2015,Andonie2022}. However, imaging data from the Chandra X-ray Observatory has shown spatially resolved and significantly extended Fe K$\alpha$ emission on scales of the order $\sim 10$s-$100$s pc in a few nearby AGN \citep[e.g.][]{Young2001,Marinucci2013,Fabbiano2017,Kawamuro2019,Ma2020,Yi2021}. 

\citet{Liu2016} used data from the Chandra High Energy Transmission Grating (HETG) to analyze the narrow Fe K$\alpha$ line in seven nearby AGN, chosen from the Chandra data archive based on the requirement of having a measurable neutral iron line in their second and third order spectra. The second and third order HETG spectra have higher spectral resolution, but effective areas that are approximately 15 times lower than the first order effective area at 6.4~keV, the energy of the neutral iron line in question \citep{Canizares2005}. Thus, using higher order spectra requires bright sources and significantly longer exposures to have a useful signal-to-noise. For each of the objects studied, \citet{Liu2016} found that the narrow Fe K$\alpha$ line width was consistently larger in the first order spectrum than in the combined second and third order spectrum. The author attributed this unexpected discrepancy to the first order spectra being overestimated due to miscalibration of the HETG. In response to the \citet{Liu2016} claim, \citet{Marshall2017} noted that the larger first order line widths were likely a result of spatially extended Fe K$\alpha$ emission in these nearby sources.

For a simple Gaussian emission line, there are three key factors which can affect the observed line width in dispersive spectroscopy, the intrinsic instrument line spread function ($\sigma_i$), the Doppler effect ($\sigma_v$), and spatial extent ($\sigma_\theta$). Assuming these all have Gaussian profiles, the total broadening of the line in detector space for a simple linear diffraction grating is given by 
\begin{equation} \label{eq:detectorwidth}
    \sigma_x^2 = \sigma_i^2 + \left(\frac{Rm\lambda}{Pc}\right)^2\sigma_v^2 + F^2\sigma_\theta^2,
\end{equation}
where $m$ is the spectral order, $R$ is the Rowland distance of the HETG, $P$ is the grating period, and $F$ is the focal length of the High Resolution Mirror Assembly \citep[HRMA; see][]{Marshall2017}. The dependence on the spectral order arises from the application of the one-dimensional grating equation, which describes the dispersion angle, $\beta$, by 
\begin{equation} \label{eq:grating}
    \sin \beta = \frac{m \lambda}{P} \approx \frac{x}{R}.
\end{equation}
Thus, spatial extent is important if the quantity $m\sigma_v$ is small and can only be ignored if $\sigma_v \gg 3400 \frac{\sigma_\theta}{m \lambda}$ km s$^{-1}$, where $\sigma_\theta$ is in arcseconds and $\lambda$ is in \AA \, \citep{Marshall2017}. Thus, at lower intrinsic velocities, the first order data will be more sensitive to spatial extent than the second and third order data. This suggests that the effect seen in \citet{Liu2016} is the result of spatial extent contributing to first order line widths.

However, actually measuring this spatial extent directly is difficult with current X-ray imaging studies. The best spatial resolution at X-ray wavelengths is $\sim 0.5"$ with Chandra. Hence, unlike with recent advances in the sub-millimeter regime allowing for the central few parsecs around nearby AGN to be resolved and mapped in detail, X-ray emission from the smallest scales around the central engine is unresolved. Furthermore, when used in spectral imaging mode, bright sources which enable a detailed study of the neutral iron line at 6.4~keV will cause significant pile-up on the Chandra detectors, in which multiple photons are counted as a single event. Pile-up causes distortion of the CCD-determined spectrum and widening of the point spread function (PSF) of the image. Thus, in practice, it is difficult to perform direct imaging studies of extended Fe K$\alpha$ emission with Chandra on scales less than a few arcseconds. 

In this work, we utilize the differences in the narrow Fe K$\alpha$ line widths for different spectral orders to obtain spatial information about the X-ray fluorescent gas around nearby AGN on scales smaller than can be probed with X-ray imaging studies. In Section \ref{sec:methods}, we outline our data reduction and analysis techniques for the sample of AGN. This includes a discussion of how the methodology was calibrated using simulated Chandra observations with MARX. In Section \ref{sec:results}, we present the results of the fits to all three spectral orders and calculate distances to the black holes, cold gas masses, and mass outflow rates. In Section \ref{sec:discussion}, we discuss the implications of our findings, including two possible explanations for discrepancies between the radius of emission measured with spatial extents and velocity widths. Finally, our results are summarized in Section \ref{sec:conclusion}. Throughout this paper, we adopt $H_0 = 70$ km s$^{-1}$ Mpc$^{-1}$, $\Omega_\Lambda = 0.73$, and $\Omega_M = 0.27$. All quoted errors correspond to the 90\% confidence level, unless otherwise noted.


\section{Methods} \label{sec:methods}

\subsection{Observations and Data Reduction} \label{subsec:data}

The Chandra HETG consists of two grating arrays, the high energy grating (HEG) for 0.8-10.0~keV and medium energy grating (MEG) for 0.4-5.0~keV. Observations with the HETG are done in combination with the Advanced CCD Imaging Spectrometer (ACIS), which can determine the energy of detected photons. The CCD-determined energy from ACIS-S allows for order sorting to be performed to obtain high-resolution HETG spectra of multiple orders.

For this work, we analyzed the same sample of seven objects from \citet{Liu2016}, NGC 1068, NGC 3783, NGC 4151, NGC 4388, NGC 4507, Mrk 3, and the Circinus galaxy. These are the AGN in the Chandra data archive which have sufficient HETG spectral data to probe the line width of the Fe K$\alpha$ line in second and third order spectra. While the second and third order spectra have higher resolution than the first order spectrum, their effective areas are smaller by a factor of approximately $1/15$ \citep{Canizares2005}. Thus, requiring high signal-to-noise second and third order spectra limits the sample to bright, nearby AGN that have been extensively observed with the Chandra HETG. Table \ref{tab:obs_info} summarizes the Chandra Observation IDs (ObsIDs) used for each source in this analysis. For Circinus, we do not include ObsID 374 in our analysis as it was approximately 7 ks long, contributing negligibly to the total observation time and lacking sufficient data above the iron line to properly constrain the fit. For NGC 4151, we only utilize the first and second order data due to issues identified in the third order data across different observations. Given that NGC 4151 has one of the strongest narrow Fe K$\alpha$ features and we only need two spectral widths to fully classify the velocity width, $\sigma_v$ and spatial extent, $\sigma_\theta$, we only use the first and second order data in this object to avoid issues with the third order spectrum.

HETG data were obtained from the Chandra archive and reprocessed using \textsc{CIAO}v4.11 \citep{Fruscione2006} and CALDBv4.8.5. We followed the standard data reduction process for grating data with one exception. We decreased the width of the masks on the grating arms used to extract the spectra to 18 pixels (approximately half the default value), which decreases the overlap between the HEG and MEG arms and hence allows us to extend our HEG analysis to higher energies. First, second, and third order spectra were extracted from each observation, and the positive and negative spectra for each order were combined to increase signal-to-noise using \texttt{combine\_grating\_spectra}. These combined spectra were then grouped to have a minimum of 1 count per bin using the \texttt{grppha} tool from HEASOFTv6.26.1. This binning is required in order to correctly employ use of the $C$ statistic during spectral fitting. Spectra were analyzed with the X-ray spectral-fitting package, XSPECv12.10.1f \citep{Arnaud1996}. Fits were completed using only HEG data, as it has a larger effective area and better resolution at 6.4~keV, the energy of the neutral Fe K$\alpha$ line under investigation. All fits were performed in the  4-9~keV range and with $C$ statistic used for minimization, which is suitable for Poisson-distributed data \citep{Cash1979}.

\begin{deluxetable*}{c c c c c c}
	\caption{Source observation information} \label{tab:obs_info}
    \tablehead{\colhead{Object} & \colhead{$z$} & \colhead{$D$} & \colhead{Ref.}\tablenotemark{b} & \colhead{ObsIDs Used} & \colhead{Total Exposure Time} \\ 
    \colhead{} & \colhead{} & \colhead{(Mpc)} & \colhead{} & \colhead{} & \colhead{(ksec)}} 
	\startdata
	NGC 1068 & 0.003793 & 14.4 & 1 & 332, 9148, 9149, 9150, 10815, 10816, 10817,  & 438.8 \\
	& & & & 10823, 10829, 10830 & \\
	NGC 3783 & 0.00973 & 38.5 & 2 & 2090, 2091, 2092, 2093, 2094, 14991, 15626, & 1145.7 \\
	& & & & 18192, 19694 &  \\
	NGC 4151 & 0.00332 & 16.1\tablenotemark{a} & 3 & 335, 3052, 3480, 7829, 7830, 16089, 16090 & 630.7 \\
	NGC 4388 & 0.00842 & 18.1 & 4 & 9276, 9277 & 268.9 \\
	NGC 4507 & 0.0118 & 50.5 & --  & 2150 & 138.2 \\
	Mrk 3 & 0.013509 & 57.9 & --  & 873, 12874, 12875, 13254, 13261, 13263, 13264, & 389.3 \\
	& & & & 13406, 14331 & \\
	Circinus & 0.001448 & 4.2 & 5 & 4770, 4771, 10223, 10224, 10225, 10226, 10832, 10833, & 660.2 \\
	& & & & 10842, 10843, 10844, 10850, 10872, 10873, 62877 &  \\
    \enddata
	\flushleft
	\tablenotetext{a}{This is the most recent distance estimate for NGC 4151, obtained using Cepheid period-luminosity measurements \citep[see ][for the details of this measurement and a discussion on the distance uncertainty for NGC 4151]{Yuan2020} }
	\tablenotetext{b}{Reference for distance measurement in Column 3. Dashes indicate that the distance is calculated from the Hubble flow ($D = cz / H_0$).}
	\tablerefs{(1) \citet{Tully1988}, (2) \citet{TullyFisher1988}, (3) \citet{Yuan2020}, (4) \citet{Tully2016}, (5) \citet{Karachentsev2013}}
\end{deluxetable*}

\subsection{Spectral Modeling} \label{subsec:modeling}

For each object in our sample, we model the 4-9~keV spectra with Galactic absorption, neutral absorption at the source, a power law continuum, and the neutral Fe K$\alpha$ emission line. All abundance values were adopted from \citet{Wilms2000}. Galactic absorption was modeled with \texttt{tbabs} with column densities adopted from \citet{HI4PI2016}. We chose to model absorption at the source with the \texttt{ztbabs} model, which accounts for a single neutral absorber, redshifted to the source redshift. There are objects in our sample whose absorption has been studied extensively and found to be more complex than our simple model. For example, NGC 3783 has been shown to have complex ionized absorption \citep[e.g.][]{Reynolds1997, Kaspi2002, Krongold2003, Netzer2003}. However, complex absorption in the X-ray spectrum is far more prevalent at softer energies, below about 2~keV. Since we are fitting from 4-9~keV, we do not include such complex absorption models as our fit is not sensitive to them. Our continuum power law and Fe K$\alpha$ emission line are both redshifted to the source redshifts given in Table \ref{tab:obs_info}. To model the Fe K$\alpha$ line, we use two Gaussian emission lines. The Fe K$\alpha$ transition is a doublet due to a difference in energy released based on the spin of the electron involved in the transition. The two lines, K$\alpha_1$ and K$\alpha_2$, are at energies of at 6.404~keV and 6.391~keV, respectively. The laboratory observed flux ratio between the two transitions is 2:1 \citep{Bearden1967}. Hence, we set the normalization of the K$\alpha_1$ line to be twice the K$\alpha_2$ normalization and set their widths to be the same. 

When modeling the Fe K$\alpha$ line as seen by the HETG, we write the observed width, $\sigma_{E \text{, measured}}$, directly in terms of the velocity width, $\sigma_v$, and spatial extent, $\sigma_\theta$, outlined in Equation \ref{eq:detectorwidth}. The instrumental width, $\sigma_i$, is accounted for in the data reduction pipeline via the response matrix. Thus, we can write Equation \ref{eq:detectorwidth} in terms of the observed energy width by applications of the grating equation and Doppler physics. Approximating the energy of the line to be much greater than the width yields
\begin{equation} \label{eq:energywidth}
    \sigma_{E \text{, measured}}^2 = \left(\frac{E_0}{c}\right)^2 \sigma_v^2 + \left(\frac{FPE_0^2}{hcmR}\right)^2 \sigma_\theta^2,
\end{equation}
where $E_0$ is the energy of the line and the other quantities are as defined in Equation \ref{eq:detectorwidth}. We approximate $E_0$ as the average of the K$\alpha_1$ and K$\alpha_2$ energies. The advantage of writing the model this way is that we can obtain more accurate error estimates on $\sigma_v$ and $\sigma_\theta$, which would be difficult otherwise because of how close we are to the instrument resolution. This yields a final XSPEC model of \texttt{tbabs(ztbabs(zpower + zgauss + zgauss))}. In fitting, we fix the Galactic column density, redshift, and centroid energies of the Gaussian emission lines and fit for the source column density, power law normalization, photon index, line normalization, velocity width, and spatial extent.

For each object, we fit simultaneously to first, second, and third order spectra for each observation. We allow the source absorption column density, power law normalization, power law photon index, and line normalization to vary for every ObsID. This accounts for changes in the source obscuration over time and the intrinsic variability of the source. These same values are fixed across all three spectral orders for a given observation when fitting, as the normalizations are corrected for the effective area with response files. We fix the value of $z$ for all model components to the values given in Table \ref{tab:obs_info}. Lastly, we fix $\sigma_v$ and $\sigma_\theta$ to be the same for all observations and spectral orders of a given source. The difference in Fe K$\alpha$ width that we are measuring then is accounted for in the dependence on the spectral order, $m$, in the spatial extent in Equation \ref{eq:energywidth}. For a constant $\sigma_v$ and $\sigma_\theta$, the velocity width contributes a constant amount across all orders, but the spatial width contribution decreases with increasing spectral order. This behavior should give the observed behavior of decreasing total line width with spectral order from \citet{Liu2016}, and permit independent measurements of $\sigma_v$ and $\sigma_\theta$. 

\subsection{Calibration with MARX} \label{subsec:marx}

\begin{figure*}[t!]
    \includegraphics[width=\textwidth]{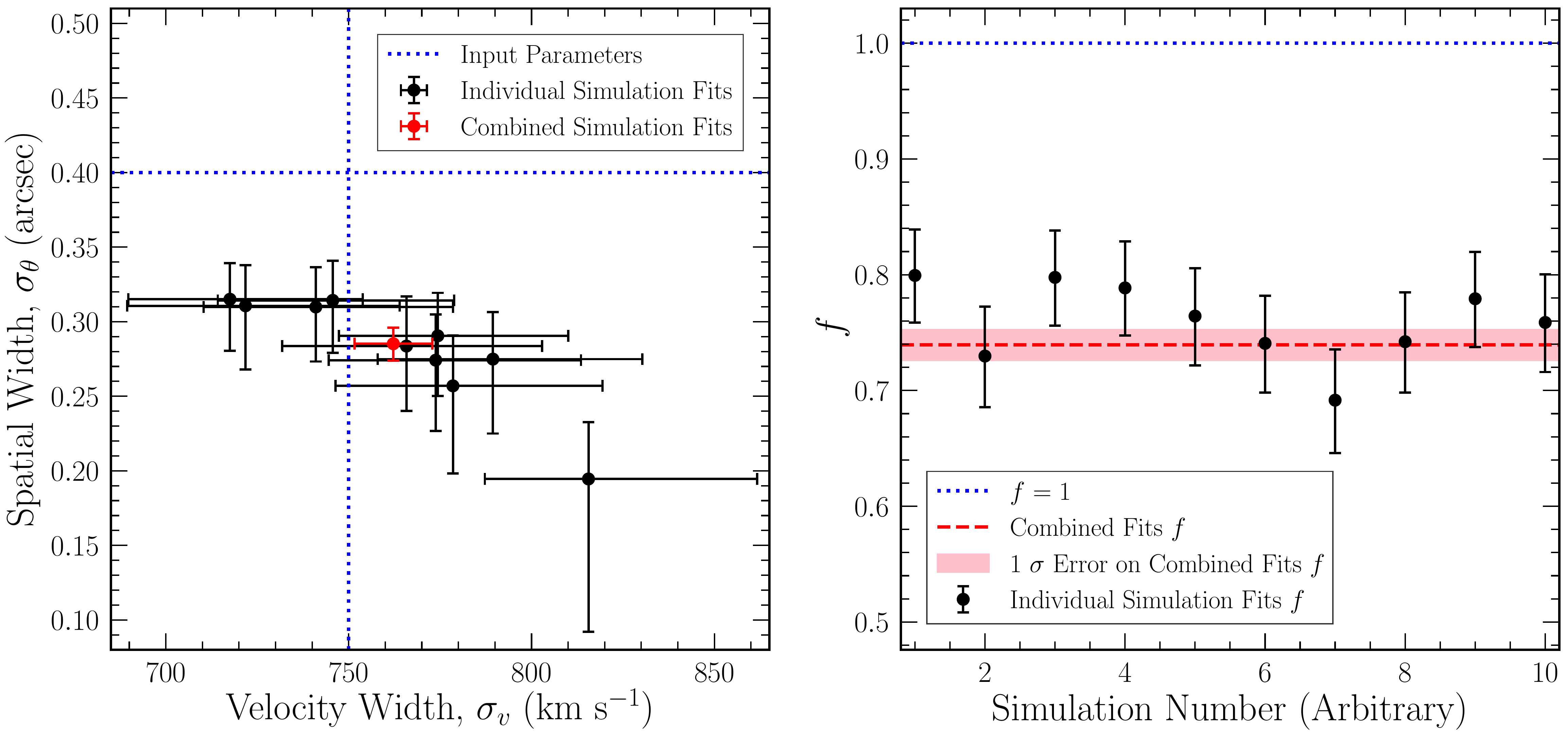}
    \caption{{\em Left}: Simulation results for 10 individual simulations with $\sigma_v = 750$ km s$^{-1}$ and $\sigma_\theta = 0.4$". Blue dotted lines show the input values of $\sigma_v$ and $\sigma_\theta$. Black data points show the fit values for $\sigma_v$ and $\sigma_\theta$ for each simulation. The red data point shows the fit value $\sigma_v$ and $\sigma_\theta$ for a simultaneous fit to all 10 simulations with $\sigma_v$ and $\sigma_\theta$ fixed to be the same between all simulations. All error bars show 1$\sigma$ (68\%) confidence. {\em Right}: Calibration factor fits for the same 10 individual simulations with $\sigma_v = 750$ km s$^{-1}$ and $\sigma_\theta = 0.4$" as in left figure. The blue dotted line shows $f = 1$. Black data points show the fit values for $f$ for each simulation. The red data dashed line shows the fit value for $f$ for a simultaneous fit to all 10 simulations. The shaded red area shows $1\sigma$ error on this combined fit value for $f$. All error bars show 1$\sigma$ (68\%) confidence.}
	\label{fig:marx}
\end{figure*}

We expect that the simple, one-dimensional grating equation used to obtain Equation \ref{eq:energywidth} may be insufficient to describe the complexity of the HRMA-HETG geometry in detail. Hence, we use MARX (version 5.5.0) to calibrate our model by simulating Chandra observations and fitting these simulated observations with the methodology described previously. MARX is a Monte Carlo, ray-tracing program used to create simulated Chandra observations \citep{Davis2012}. We simulated a power law source with point geometry, since the corona has been shown to be compact and quite close to the central supermassive black hole. The Fe K$\alpha$ emission however, is spatially extended. Hence, we modelled the Fe K$\alpha$ doublet with some intrinsic velocity width, $\sigma_v$, and extended the emission over a Gaussian source with width $\sigma_\theta$. The two components, the corona and the line source, were concatenated to make a single simulated observation using \texttt{marxcat} and then processed in the standard way with MARX and CIAO tools. 

Simulations were run for a range of $\sigma_v$ and $\sigma_\theta$ values, spanning a range of reasonable narrow Fe K$\alpha$ line widths, including cases where $\sigma_v$ dominates $\sigma_\theta$, $\sigma_\theta$ dominates $\sigma_v$, and the two contribute similarly. The left panel of Figure \ref{fig:marx} shows an example of the fit that uses Equation \ref{eq:energywidth} to relate the total line width in each order to $\sigma_v$ and $\sigma_\theta$ for 10 simulations using input values of $\sigma_v = 750$ km s$^{-1}$ and $\sigma_\theta = 0.4$". Each simulation was run for 10 Ms, giving us excellent data with which to test the model. Following the reprocessing of the data using standard CIAO tools, the first, second, and third order spectra were fit using the same procedure as described in Section \ref{subsec:data} and with the same model as described in Section \ref{subsec:modeling}. The 10 simulations were each fit individually and then fit simultaneously, with $\sigma_v$ and $\sigma_\theta$ fixed across all simulations, similarly to how we fit simultaneously to all observations for a given object.

The left panel of Figure \ref{fig:marx} shows that while the model fits the velocity width accurately, the use of Equation \ref{eq:energywidth} leads to a spatial width that is significantly underestimated. This trend also holds for the other combinations of $\sigma_v$ and $\sigma_\theta$ that we tested. It is thus apparent that Equation \ref{eq:energywidth} does not fully capture the response of the HETG to finite spatial extent. We adopt a calibration factor, $f$, to attain the true value of $\sigma_\theta$ when fitting. This is implemented by inserting this calibration factor $f$ into the spatial width term of Equation \ref{eq:energywidth}, giving
\begin{equation} \label{eq:calibrating}
    \sigma_{E \text{, measured}}^2 = \left(\frac{E_0}{c}\right)^2 \sigma_v^2 + \left(f\frac{FPE_0^2}{hcmR}\right)^2 \sigma_\theta^2.
\end{equation}
The value of $f$ is then determined by fixing $\sigma_v$ and $\sigma_\theta$ to the input values and fitting the simulated data to $f$. The right panel of Figure \ref{fig:marx} shows the result of fitting to $f$ for the same simulations as in the left panel. We adopt the average value of the simultaneous fits for five different input combinations of $\sigma_v$ and $\sigma_\theta$, $f = 0.73$, for the analysis of the data.

We stress that this is a first order correction to Equation \ref{eq:energywidth}. While the calibration factors show general consistency across a range of velocity and spatial widths, there is some hint of small deviations across different input values of $\sigma_v$ and $\sigma_\theta$. We suspect that there is a second order correction due to a deviation from gaussianity that is on the order of 10\% of $f$ and potentially dependent on $\sigma_v$, $\sigma_\theta$, and the signal-to-noise. For the purposes of this work, we simply adopt the first order correction, $f = 0.73$, as these small, second order corrections are a negligible source of uncertainty.

\subsection{MCMC Fitting of AGN Data} \label{subsec:mcmc}

Adopting a calibration factor of $f = 0.73$, we performed fits to the spectral model described in Section \ref{subsec:modeling} using the Markov Chain Monte Carlo (MCMC) method implemented in XSPEC. We utilized the Goodman-Weare algorithm \citep{Goodman2010}, with 5 independent chains for each object, where each chain has $10^6$ steps and 500 walkers. For NGC 3783, NGC 4151, NGC 4388, NGC 4507, and Circinus, the MCMC chains were started based on the covariance matrix derived from the initial fit to the data. In the Circinus chains, the covariance matrix used to populate the initial distribution of walkers was scaled by $10^{-1}$ to allow the walkers to fully explore parameter space. The burn-in length for these objects was $2 \times 10^5$ chain steps. For NGC 1068 and Mrk 3, the MCMC chains were populated from a uniform distribution for all parameters across a wide range of reasonable values. This was done because the initial fit performed in XSPEC using $C$ statistic minimization collapsed to essentially zero spatial extent, resulting in subsequent chains not fully exploring $\left(\sigma_v, \sigma_\theta\right)$ parameter space. These objects required a longer burn-in length of $10^6$ chain steps for convergence. For each object, all 5 chains are then combined into a single composite chain with a total of $5 \times 10^6$ chain steps and used for analysis. Reported best fit values are median values from the combined MCMC chains and uncertainties are the 5th and 95th percentile chain values.

For each object, we checked convergence with the Geweke diagnostic \citep{Geweke1992} and the Gelman-Rubin scale reduction factor, $\hat{R}$ \citep{Gelman1992}. The Geweke diagnostic, which compares the mean of a given parameter at the beginning of the chain (first 10\%) and at the end of the chain (last 50\%), was found to be $\left|z_G\right| < 1$ for each fit parameter in each chain. The Gelman-Rubin scale reduction factor measures convergence across independent chains, and satisfies $\hat{R} \lesssim 1.2$ for each fit parameter. Together, these two convergence factors indicate that the chains have converged to the posterior distribution.


\section{Results} \label{sec:results}

\subsection{Velocity and Spatial Widths}

For each object in the sample, we obtained a fit to $\sigma_v$ and $\sigma_{\theta}$ with uncertainty determined using the MCMC posterior distributions. These fit values are given in Table \ref{tab:sigv_sigth}, along with the line width in energy space for each order as calculated with Equation \ref{eq:calibrating}. The values and errors on the line width in each order are computed directly from the MCMC chains, taking $\sigma_v$ and $\sigma_\theta$ for each chain step and converting these to a total energy width for each order. We find that the unexpected behavior from \citet{Liu2016}, in which first order line widths are larger than their second and third order counterparts, can indeed be described by the intrinsic velocity width and spatial extent, as proposed in \citet{Marshall2017}.

\begin{deluxetable*}{c c c c c c c c c c c}
	
	\caption{Fits to $\sigma_v$ and $\sigma_{\theta}$, resulting Fe K$\alpha$ line widths for different order spectra, and Fe K$\alpha$ equivalent widths \label{tab:sigv_sigth}}
	
	\tablehead{\colhead{Object} & \colhead{$\sigma_v$} & \colhead{$v_\mathrm{FWHM}$} & \colhead{$r_v$} & \colhead{$\sigma_\theta$} & \colhead{$r_\theta$} & \colhead{$\sigma (\pm 1)$} & \colhead{$\sigma (\pm 2)$} & \colhead{$\sigma(\pm 3)$} & \colhead{EW$_\mathrm{unabs}$} & \colhead{$C$ / dof} \\
	& \colhead{(km s$^{-1}$)} & \colhead{(km s$^{-1}$)} & \colhead{(pc)} & \colhead{(arcsec)} & \colhead{(pc)} & \colhead{(eV)} & \colhead{(eV)} & \colhead{(eV)} & \colhead{(eV)} & }		
	
	\startdata
	NGC 1068 & $898_{-432}^{+231}$ & $2110_{-1020}^{+540}$ & $0.013_{-0.005}^{+0.035}$ & $0.31_{-0.28}^{+0.41}$ & $21.8_{-19.6}^{+28.4}$ & $21.6_{-3.6}^{+4.0}$ & $19.7_{-6.1}^{+4.6}$ & $19.4_{-7.7}^{+4.8}$ & 10 - 510 & $2189.7/2617$ \\
    NGC 3783 & $592_{-245}^{+177}$ & $1390_{-580}^{+420}$ & $0.088_{-0.036}^{+0.168}$ & $0.30_{-0.26}^{+0.25}$ & $55.5_{-49.0}^{+46.0}$ & $15.5_{-2.7}^{+2.9}$ & $13.3_{-3.4}^{+3.3}$ & $12.9_{-4.2}^{+3.6}$ & $79.5_{-23.9}^{+29.8}$ & $12331.4/13673$ \\
    NGC 4151 & $894_{-359}^{+185}$ & $2100_{-850}^{+440}$ & $0.059_{-0.019}^{+0.106}$ & $0.39_{-0.34}^{+0.33}$ & $30.3_{-26.6}^{+25.5}$ & $22.1_{-2.3}^{+2.4}$ & $19.8_{-5.0}^{+3.4}$ & $19.4_{-6.4}^{+3.7}$ & $108.6_{-57.8}^{+55.2}$ & $7925.8/8501$ \\
    NGC 4388 & $619_{-506}^{+413}$ & $1460_{-1190}^{+970}$ & $0.023_{-0.015}^{+0.672}$ & $0.57_{-0.49}^{+0.33}$ & $49.9_{-42.8}^{+29.2}$ & $21.1_{-5.3}^{+5.8}$ & $15.5_{-5.2}^{+7.0}$ & $14.2_{-6.7}^{+8.0}$ & $101.3_{-19.7}^{+20.4}$ & $2087.7/2367$ \\
    NGC 4507 & $561_{-458}^{+379}$ & $1320_{-1080}^{+890}$ & $0.806_{-0.519}^{+22.844}$ & $0.36_{-0.33}^{+0.38}$ & $89.1_{-79.6}^{+94.1}$ & $16.7_{-6.0}^{+7.2}$ & $13.3_{-5.8}^{+7.3}$ & $12.5_{-6.8}^{+7.7}$ & $45.5_{-9.6}^{+10.8}$ & $1091.2/1117$ \\
    Mrk 3 & $976_{-556}^{+259}$ & $2300_{-1310}^{+610}$ & $0.485_{-0.182}^{+2.142}$ & $0.36_{-0.33}^{+0.45}$ & $101.0_{-91.5}^{+126.3}$ & $23.7_{-3.9}^{+4.4}$ & $21.4_{-7.5}^{+5.2}$ & $21.1_{-9.7}^{+5.4}$ & $157.9_{-81.4}^{+103.8}$ & $2138.7/2824$ \\
    Circinus & $341_{-108}^{+106}$ & $800_{-260}^{+250}$ & $0.015_{-0.006}^{+0.017}$ & $0.37_{-0.10}^{+0.07}$ & $7.5_{-2.0}^{+1.4}$ & $12.4_{-1.1}^{+1.1}$ & $8.9_{-1.3}^{+1.5}$ & $8.0_{-1.7}^{+1.9}$ & 20 - 770 & $6452.9/7400$ \\
	\enddata
	
	\flushleft
	\tablecomments{Columns are: (1) Object name, (2) Measured velocity width, (3) Full width-half maximum (FWHM) velocity, (4) Distance from the SMBH measured from the FWHM velocity (Equation \ref{eq:r_v}), (5) Measured spatial extent, (6) Distance from the SMBH measured from the spatial extent (Equation \ref{eq:r_theta}), (7) First order Fe K$\alpha$ line width, (8) Second order Fe K$\alpha$ line width, (9) Third order Fe K$\alpha$ line width, (10) Fe K$\alpha$ equivalent width from first order spectra corrected for absorption and averaged over all observations, (11) Average value of the $C$ statistic from MCMC chains and degrees of freedom}

\end{deluxetable*}

\begin{figure*}[t!]
    \centering
    \includegraphics[width=\textwidth]{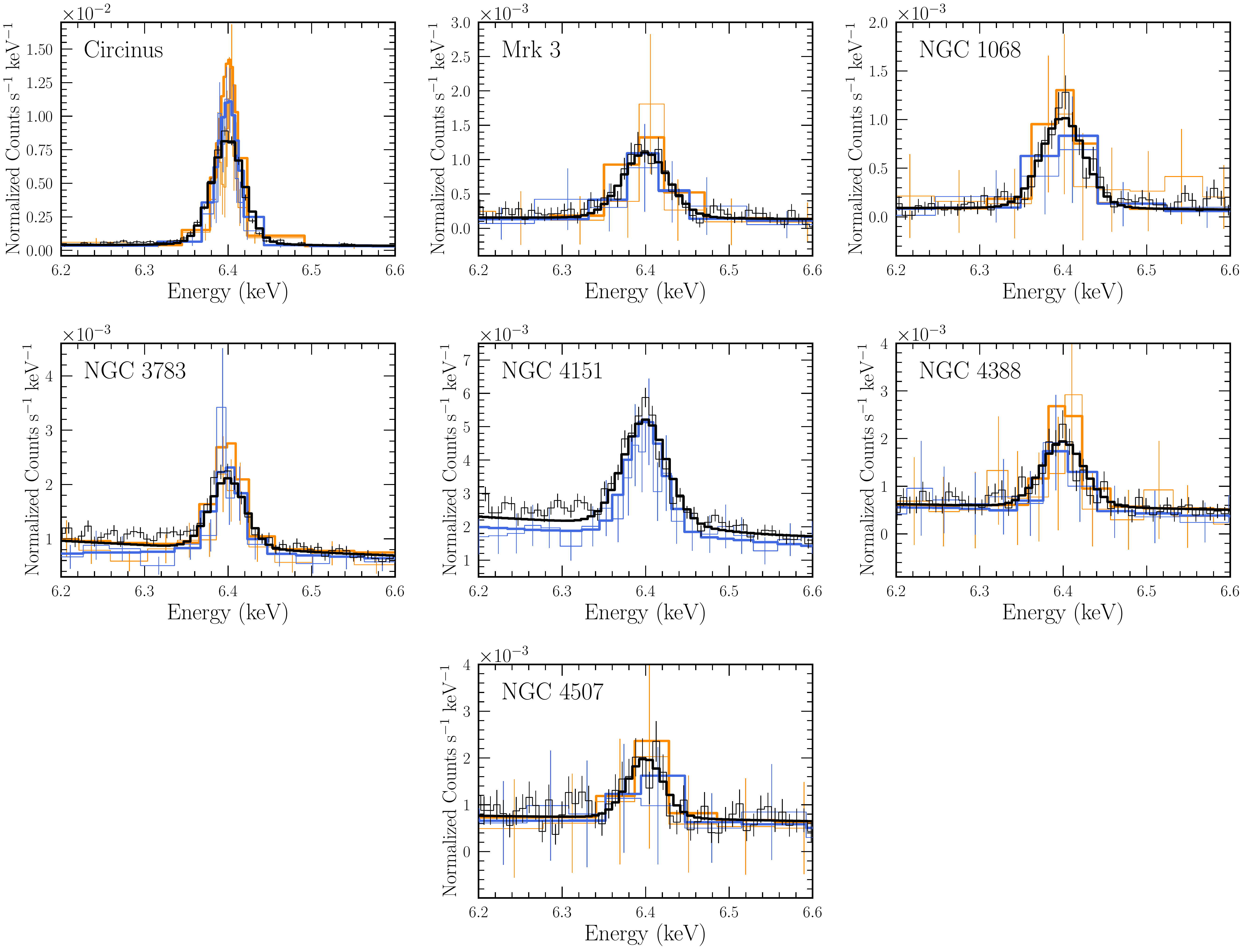}
    \caption{Spectra for all seven objects in our sample with all ObsIDs combined together and corrected for redshift. Black lines with error bars indicate the first order spectrum, decreased by a factor of 15 to roughly account for the difference in effective areas. Blue and orange lines with error bars indicate the second and third order spectra, respectively. For NGC 4151, only the first and second order spectra are shown. All first order spectra are binned to a minimum signal-to-noise of 3. For NGC 3783 and NGC 4151, the higher order spectra are also binned to a minimum signal-to-noise of 3. For Circinus, the higher order spectra are binned to a minimum signal-to-noise of 2. Due to low signal-to-noise in the remaining objects, the higher order spectra in the remaining objects are binned in groups of 5 bins. All binning is done solely for visual purposes in plotting. The models with fit parameters from Tables \ref{tab:sigv_sigth} and \ref{tab:fitpars} averaged over all observations are over-plotted in bold.}
	\label{fig:spectra}
\end{figure*}

\vspace{-0.8cm}
First, second, and third order spectra for all seven objects in our sample are shown in Figure \ref{fig:spectra}. These spectra contain the combined data from all observations listed in Table \ref{tab:obs_info}. Each spectrum is plotted with error bars, and the model for each order containing fit parameters averaged over all observations is over-plotted in bold. The first order spectrum is reduced in normalization by a factor of 15 to roughly account for the difference in effective areas between the orders, and all spectra are binned for plotting purposes only. However, we stress that this is not how we performed the fits, as we fit each observation for each object to its own absorption, continuum parameters, and line normalization to account for variability (see Section \ref{subsec:modeling} for details). This is simply a representation of the data we are working with and the features of the spectrum at 6.4~keV.

We note that our errors on the second and third order line widths are significantly smaller than those in \citet{Liu2016}, which show that the measured line width in the combined higher order spectrum is an upper limit for four of the seven sources. Our errors are smaller because we did not fit the second and third order data individually. Rather, we fit all three orders simultaneously and linked them through writing the line widths in terms of $\sigma_v$ and $\sigma_{\theta}$ and tying these values across all three orders. Given that the first order data has significantly higher signal-to-noise than the second and third order data, our fit to $\sigma_v$ and $\sigma_{\theta}$ is most sensitive to first order data. The second and third order data allows us to break the degeneracy between $\sigma_v$ and $\sigma_{\theta}$ and estimate where in this parameter space each object lies.

\begin{figure*}[t!]
    \centering
    \includegraphics[width=\textwidth]{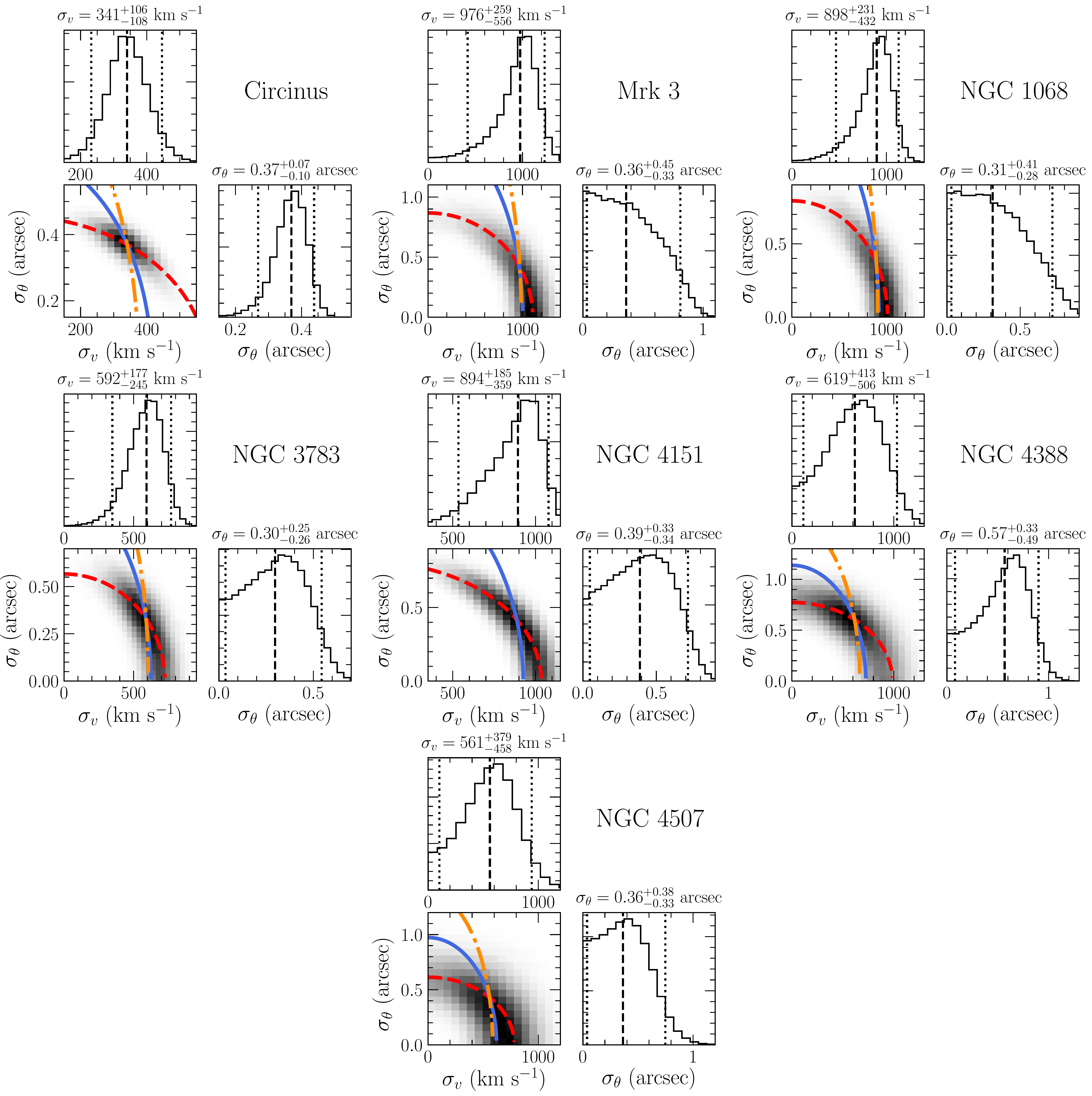}
    \caption{Corner plots for each object in the sample. For each corner plot, the two histogram plots show the posterior distributions for $\sigma_v$ and $\sigma_{\theta}$. The dashed vertical line shows the median chain value, reported in Table \ref{tab:sigv_sigth}. The two dotted vertical lines show the 90\% confidence bounds on the velocity widths and spatial extents. The other plot shows the density of chain steps in $(\sigma_v, \sigma_{\theta})$ space. The lines show lines of constant line width, $\sigma_{E \text{, measured}}$, in each order at the median values given in Table \ref{tab:sigv_sigth}. The red dashed line corresponds to first order, blue solid line to second order, and orange dot-dashed line to third order.}
	\label{fig:corner_all}
\end{figure*}

However, we also see less spatial extent in NGC 1068 and Mrk 3 than would be expected given the fits from \citet{Liu2016}. That is, our second and third order line widths for these objects are significantly larger than those from \citet{Liu2016}, which are quite small (upper limits of $\sim 5-7$ eV).  We investigate this discrepancy by fitting the second and third order data of both of these objects individually and fixing the continuum parameters to those from our overall fit. In the third order spectrum of NGC 1068, we find anomalously high and poorly constrained energy widths on the order of $\sim 100-400$ eV, a consequence of the limited data quality in the third order spectrum. In the second order spectrum for NGC 1068 and in both second and third order spectra for Mrk 3, we find larger line widths than those from \citet{Liu2016}, on the order of $\sim 25-50$ eV, albeit it with quite large error bars ranging from $\sim 10-25$ eV. With slightly different fitting methods and more recent calibration files implemented, we find no evidence for such small line widths in higher order HETG data for NGC 1068 and Mrk 3.

Corner plots for each object in our sample are shown in Figure \ref{fig:corner_all}. In each corner plot, the two histograms show the posterior distributions of $\sigma_v$ and $\sigma_{\theta}$. The bottom left plot shows the density of chain steps in $(\sigma_v, \sigma_{\theta})$ space. Lines of constant total line width are over-plotted, where the energy is equal to the values given in Table \ref{tab:sigv_sigth}, the median $\sigma_\mathrm{E, \, measured}$ from Equation \ref{eq:calibrating}. The red dashed line, which shows constant first order line width, tends to trace the shape of the distribution of $(\sigma_v, \sigma_{\theta})$ values quite well. This illustrates that our fitting is most sensitive to first order data and is picking out a roughly constant first order line width over the entire chain. The blue and green dashed lines show constant line width in second and third order, respectively. These values are breaking the degeneracy between $\sigma_v$ and $\sigma_\theta$ that would not be possible with first order data alone. From the plots, we can see that with increasing order, the effect of $\sigma_{\theta}$ decreases. In third order, the green dashed line is nearly vertical, indicating that the third order data is relatively insensitive to the value of $\sigma_{\theta}$ compared to the first order data.

For each object and each observation, additional fit parameters from the MCMC analysis are given in Section \ref{sec:append_tab} of the Appendix. These include the source absorption column density ($N_H$), photon index ($\Gamma$), power law normalization ($K_\mathrm{power}$), and the line normalization ($K_\mathrm{line}$). We note that for NGC 1068 and Circinus, our source absorption column densities are significantly lower than those reported in the literature from more complex modeling of the absorption \citep[e.g.][]{Matt1997,Matt1999,Arevalo2014,Bauer2015,Zaino2020}. These two sources are known Compton-thick AGN, whereby the obscuring material is optically thick to electron scattering (with $N_H \geq 1.5 \times 10^{24}$~cm$^{-2}$). Similarly, Mrk 3 has been shown to be near the Compton-thick limit along the line of sight \citep[e.g.][]{Cappi1999,Yaqoob2015,Guainazzi2016}. When the nucleus is this obscured, it can be difficult to obtain information about the intrinsic X-ray properties. This is addressed further in Section \ref{subsec:comptonthick}, where we discuss how this impacts our results and outline prior work that has been done in an attempt to unveil the nuclei of these objects and determine their intrinsic X-ray properties.

\subsection{Extended Fe K$\alpha$ Emission} \label{subsec:extended}

From both $\sigma_v$ and $\sigma_{\theta}$, we can estimate the distance of the gas from the black hole. Assuming the motion is purely gravitational with the potential dominated by the supermassive black hole, and the emitting gas is far enough away from the black hole to ignore general relativistic effects, this is given by the simple Keplerian formula, 
$r = G M_\mathrm{BH} / \langle v^2 \rangle$.
Adopting the assumption that the velocity dispersion is related to the FWHM velocity by $\langle v^2 \rangle = \frac{3}{4}v_\mathrm{FWHM}^2$ \citep{Netzer1990}, the distance from the black hole is 
\begin{equation} \label{eq:r_v}
    r_v = \frac{4GM_\mathrm{BH}}{3v_\mathrm{FWHM}^2}.
\end{equation}
We adopt black hole masses from the literature, given in Table \ref{tab:mass}. For our fit values to $\sigma_v$, this gives distances on the order of $r_v \sim 0.01-1$ pc, or emission coming from the BLR or inner parts of the torus. 

The spatial extent, $\sigma_{\theta}$, provides another measurement for the distance of the material from the black hole. With the distance to the galaxy, $D$, the spatial extent in angular units is converted to a physical distance simply by
\begin{equation} \label{eq:r_theta}
    r_{\theta} = D \sigma_{\theta}.
\end{equation}
The two most distant galaxies, NGC 4507 and Mrk 3, can be well-described by Hubble flow distances, $D = cz / H_0$. For the remainder of the galaxies in our sample, we use distances determined from other measures in the literature, given in Table \ref{tab:obs_info}. For our fits to $\sigma_{\theta}$, we obtain distances on the order of $r_\theta \sim 5-100$ pc, indicating that the emission is more extended than the typical BLR or inner torus. 

\begin{figure}[t!]
    \centering
    \includegraphics[width=\columnwidth]{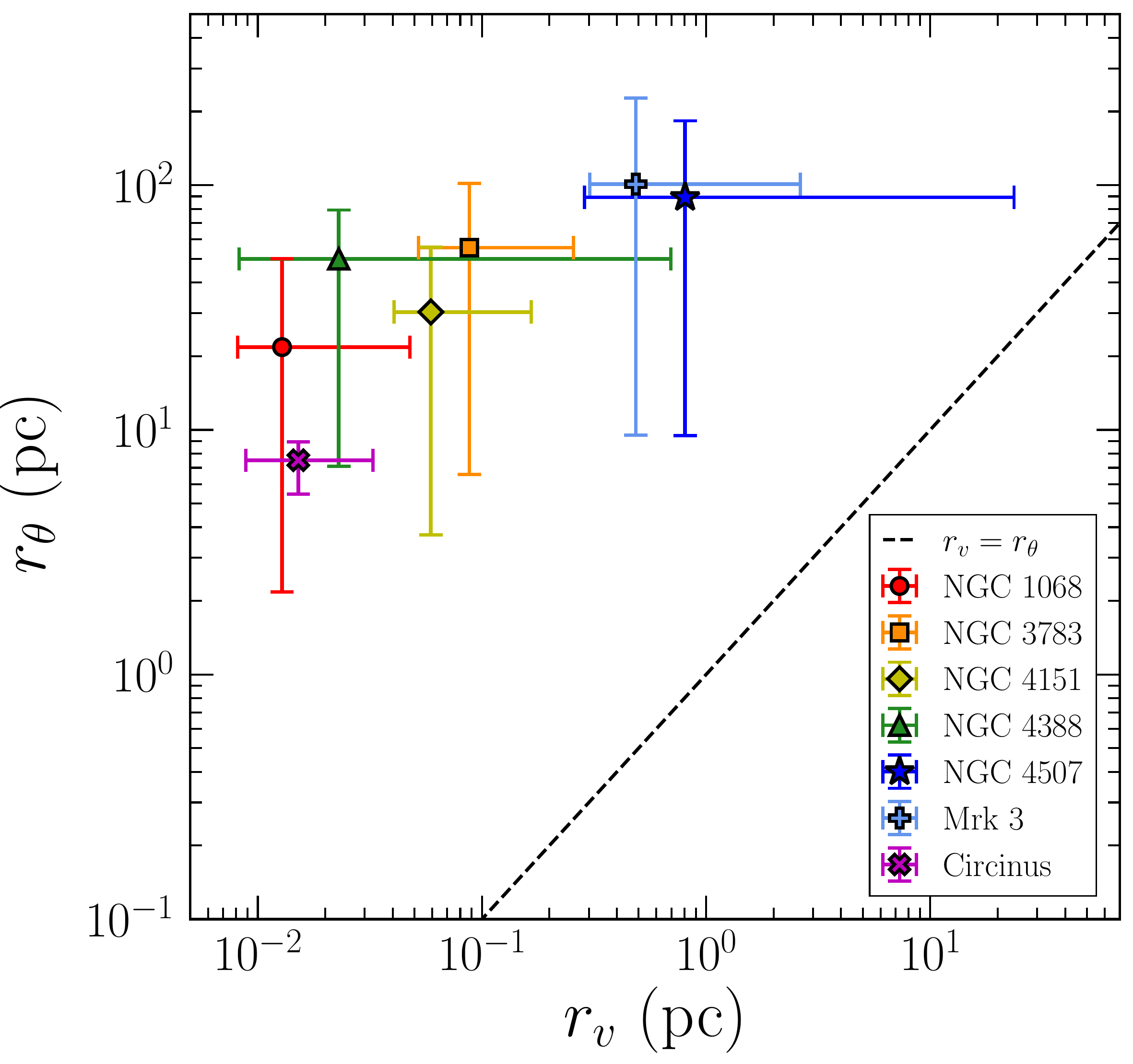}
    \caption{Comparison of the distance of the emitting gas from the black hole, using two different measurements. The black dashed line shows equality between the two estimates. The distances measured by the spatial extent, $r_\theta$, are consistently larger than the distances than the distance measured by the velocity width, $r_v$ under the assumption of gravitational motions for all seven objects in the sample. The values in this figure are reported in Table \ref{tab:sigv_sigth} and two possible explanations for the discrepancy are discussed in Section \ref{subsec:origin}.}
	\label{fig:rv_vs_rtheta}
\end{figure}

These distances for each object are given in Table \ref{tab:sigv_sigth} and a comparison of the distances measured using the spatial extent and using the velocity width is shown in Figure \ref{fig:rv_vs_rtheta}. We find that distances measured with the velocity width are significantly smaller than the distances from spatial extent. Further discussion of how this informs our understanding of the origin of the narrow Fe K$\alpha$ line follows in Section \ref{subsec:origin}.

\subsection{Mass Estimates} \label{subsec:mass}

Following the steps outlined in \citet{Hitomi2018}, which builds off of the equivalent width analysis originally presented in \citet{Reynolds2000}, we estimate the mass of the X-ray irradiated gas. The average, unabsorbed equivalent width measured from the spectral model is used to calculate the quantity $f_\mathrm{cov} N_H$ via Equation (1) of \citet{Hitomi2018},
\begin{equation} \label{eq:eqw}
    EW_\mathrm{Fe} \approx 65 \, f_\mathrm{cov} \left(\frac{Z_\mathrm{Fe}}{Z_\odot}\right) \left(\frac{N_H}{10^{23} \, \mathrm{cm}^{-2}}\right) \, \mathrm{eV}. 
\end{equation}
Here $f_\mathrm{cov}$ is the covering fraction of the material, allowing for a non-uniform, clumpy or patchy distribution of material, and $N_H$ is the column density. This calculation stems from counting fluorescing iron atoms based on the photoionization probability, fluorescent yield, and incident flux, and is scaled from the initial calculation of $EW_\mathrm{Fe, \, max} \approx 65$ eV with $Z_\mathrm{Fe} = Z_\odot$ and $N_H = 10^{23}$~cm$^{-2}$ for NGC 4258 from \citet{Reynolds2000}. The details behind this calculation are given in full in Section \ref{sec:append_eqw} of the Appendix. Then, using the quantity $f_\mathrm{cov} N_H$ and assuming the gas is located in a shell at radius $r$, the gas mass can be estimated using $M_\mathrm{gas} = 4 \pi r^2 f_\mathrm{cov} N_H \mu m_p$, where $\mu$ is the mean molecular weight. We adopt $\mu = 1.3$ for neutral, solar metallicity material, $Z_\mathrm{Fe} = Z_\odot$ for solar iron abundance, and use the distance measured from our fit to the spatial extent, $r_\theta$.

This calculation requires the total equivalent width of the Fe K$\alpha$ line, which is found by adding the equivalent widths for the K$\alpha_1$ and K$\alpha_2$ transitions from the first order spectrum and averaging over all observations for a given object. However, the equivalent width in Equation \ref{eq:eqw} is the unabsorbed equivalent width. To compute the unabsorbed equivalent width for gas mass estimates, we first fit to the model \texttt{tbabs(zgauss + zgauss + ztbabs*zpower)}, where only the continuum is being obscured at the source. Then, we remove the \texttt{ztbabs} component to compute the Fe K$\alpha$ equivalent width measured against the unabsorbed incident X-ray flux. To fit this model, we fix the fit parameters $N_H$, $\Gamma$, $K_\mathrm{power}$, $\sigma_v$ and $\sigma_\theta$, leaving only the line normalization ($K_\mathrm{line}$) free to vary as we expect this to be the only affected parameter in changing the model. This assumption was explicitly verified for NGC 3783, NGC 4151, NGC 4388, and NGC 4507. The normalization of the Fe K$\alpha$ line decreases in this new model because the line emission is no longer attenuated by the absorption. The amount of decrease in $K_\mathrm{line}$ is positively correlated with column density of the absorber, $N_H$, as expected.

These unabsorbed equivalent width values are given in Table \ref{tab:sigv_sigth}. We report the Compton-thin equivalent widths with errors that account for the uncertainties in the original fit parameters as well as those from the new fit to $K_\mathrm{line}$. To assess these errors, we took 100 random draws from the MCMC chains for the fixed fit parameters and followed the above procedure to fit to $K_\mathrm{line}$ for each draw. Then for each draw, we took 100 random draws for $K_\mathrm{line}$ from a Gaussian distribution centered at the fit value of $K_\mathrm{line}$ with standard deviation based on the fit error. This gives us a distribution of unabsorbed equivalent width values from which we compute uncertainties. 

For the two Compton-thick sources in our sample (NGC 1068 and Circinus), accurately determining the intrinsic continuum is extremely difficult as the majority of the observed emission in the 4-9~keV range is scattered and reprocessed emission. The intrinsic nuclear continuum is best determined by including hard X-ray data in the modeling, which contains more transmitted emission than the soft energy band. Such modeling is beyond the scope of this work and hence, for the Compton-thick sources in our sample, we report a range of equivalent width values, which represent the extremes of possible equivalent width values depending on different assumptions about the continuum emission.

As an upper bound on the equivalent width, we assume that the observed continuum is the continuum that the Fe K$\alpha$ emitting material sees. In reality, this is likely all scattered emission, but this provides a lower bound on the continuum flux and hence an upper bound on the equivalent width. To compute the lower bound for the equivalent width in our Compton-thick sources, we use the nuclear continuum estimated from detailed, broad-band modeling including NuSTAR data from prior work (e.g. NGC 1068--\citealt{Bauer2015,Marinucci2016}, Circinus--\citealt{Arevalo2014,Uematsu2021}). In this scenario, the Fe K$\alpha$ emitting material sees a much stronger continuum, leading to a lower equivalent width measurement. The picture is certainly more complicated than these two extremes, especially given that the Fe K$\alpha$ emitting material likely emits at a range of radii, thus seeing different illuminating radiation and being obscured differently depending on the location of the emission. However, the ranges provided in Table \ref{tab:sigv_sigth} for the equivalent width of the Compton-thick sources capture the uncertainty in determining the intrinsic continuum that the Fe K$\alpha$ emitting material sees and the location of the Fe K$\alpha$ emitting material relative to the obscurer. We further discuss the uncertainty associated with modeling Compton-thick sources in Section \ref{subsec:comptonthick}.

Adopting the unabsorbed equivalent widths from Table \ref{tab:sigv_sigth}, we compute the mass of the cold, X-ray irradiated gas. The masses and their uncertainties are given in Table \ref{tab:mass}. As with the equivalent widths, we report a range of gas masses corresponding to the equivalent width ranges for the Compton-thick sources in our sample. For the Compton-thin sources, the uncertainty in the gas masses includes the uncertainty in both the spatial extent and the unabsorbed equivalent width by using the distribution of unabsorbed equivalent width values and corresponding $r_\theta$ chain value from each equivalent width measurement. Most of our mass measurements are close to upper limits, with large uncertainty due to the large uncertainty on the spatial extent of the emitting material. It is important to also note that we assumed solar abundance for iron, as our fit does not constrain this value, but that the gas mass is proportional to $Z_\mathrm{Fe}^{-1}$. Having non-solar abundances could change the result by a factor of a few, but we expect that our results are accurate to within an order of magnitude.

In Section \ref{subsec:origin}, we explore two possibilities for the discrepancy between the distance measurements from Section \ref{subsec:extended}, one of which is that the Fe K$\alpha$ emitting material is outflowing. Using the mass measurements, we can thus estimate the mass outflow rates of the cold gas. Adopting the velocity width, $\sigma_v$, to be the characteristic outflow velocity of the gas and $r_\theta$ to be the distance of the emitting gas from the black hole, the mass outflow rate is given by
\begin{equation} \label{eq:mass_out_rate}
    \dot{M}_\mathrm{out} = \frac{M_\mathrm{gas} v_\mathrm{out}}{r} \sim \frac{M_\mathrm{gas} \sigma_v}{r_\theta}.
\end{equation} 
The outflow rates for each object is given in Table \ref{tab:mass}. We note that, like the mass estimates from earlier in the section, many of these values are highly uncertain. The similarities between these outflow rates and molecular outflow rates are discussed further in Section \ref{subsec:compareALMA}.

\begin{deluxetable*}{c c c c c c c c}
	
	\caption{Cold Gas Mass, Luminosity, and Mass Outflow Rates} \label{tab:mass}
	
	\tablehead{\colhead{Object} & \colhead{$M_\mathrm{gas}$} & \colhead{$\dot{M}_\mathrm{out}$} & \colhead{$\log L_\mathrm{bol}$} & \colhead{Ref.} & \colhead{$\log M_\mathrm{BH}$} & \colhead{Ref.} & \colhead{$\lambda_\mathrm{Edd}$} \\
	& \colhead{($10^7 M_\odot$)} & \colhead{($M_\odot$ yr$^{-1}$)} & \colhead{(erg s$^{-1}$)} & & \colhead{($M_\odot$)} & & }
	
	\startdata
    NGC 1068 & 0.06-5\tablenotemark{$\dagger$} & 20-2100\tablenotemark{$\dagger$} & $44.7$ & 1 & $7.0$ & 5 & $0.399$ \\
    NGC 3783 & $5.3_{-5.2}^{+12.7}$ & $500_{-420}^{+480}$ & $44.2$ & 2 & $7.47$ & 6 & $0.042$ \\
    NGC 4151 & $1.8_{-1.8}^{+5.3}$ & $440_{-380}^{+570}$ & $43.7$ & 3 & $7.66$ & 7 & $0.008$ \\
    NGC 4388 & $4.9_{-4.7}^{+8.7}$ & $460_{-400}^{+620}$ & $43.7$ & 2 & $6.93$ & 8 & $0.047$ \\
    NGC 4507 & $6.3_{-6.2}^{+23.8}$ & $360_{-300}^{+600}$ & $44.3$ & 2 & $8.39$ & 9 & $0.006$ \\
    Mrk 3 & $33.8_{-33.6}^{+135.9}$\tablenotemark{$\dagger$} & $2870_{-2530}^{+3940}$\tablenotemark{$\dagger$} & $44.8$ & 2 & $8.65$ & 10 & $0.011$ \\
    Circinus & 0.02-0.9\tablenotemark{$\dagger$} & 10-410\tablenotemark{$\dagger$} & $43.6$ & 4 & $6.23$ & 11 & $0.186$ \\
	\enddata
	
	\flushleft
	\tablecomments{Columns are: (1) Object name, (2) Mass of Fe K$\alpha$ emitting gas, (3) Mass outflow rate, calculated using Equation \ref{eq:mass_out_rate}, (4) Bolometric luminosity, (5) Reference for bolometric luminosity, (6) Black hole mass, (7) Reference for black hole mass, (8) Eddington ratio ($\lambda_\mathrm{Edd} = L_\mathrm{bol} / L_\mathrm{Edd}$, where $L_\mathrm{Edd} = 1.26 \times 10^{38} \left(M_\mathrm{BH}/M_\odot\right)$ erg s$^{-1}$)}
	\tablenotetext{\dagger}{See Sections \ref{subsec:mass} and \ref{subsec:comptonthick} for discussions on how our modeling impacts these measurements in sources which are close to or exceed the Compton-thick limit}
	
	\tablerefs{(1) \citet{Lopez-Rodriguez2018}, (2) \citet{Vasudevan2010}, (3) \citet{Duras2020}, (4) \citet{Moorwood1996}, (5) \citet{Greenhill1996}, (6) \citet{Peterson2004}, (7) \citet{Bentz2006}, (8) \citet{Kuo2011}, (9) \citet{Winter2009}, (10) \citet{Woo2002}, (11) \citet{Greenhill2003}}
	
\end{deluxetable*}

\subsection{Trends with AGN Properties} \label{subsec:trends}

We investigated trends between our results for $\sigma_v$ and $\sigma_\theta$ and AGN properties such as X-ray luminosity, obscuration, and Eddington ratio. However, due to the large uncertainties on all objects except for Circinus and our relatively small sample size, we are unable to find any significant trends with AGN properties. Likewise, we are not able to draw any firm conclusions on the lack of correlations, but suggest that this is best left as work that could be investigated with future X-ray missions, such as XRISM.


\section{Discussion} \label{sec:discussion}

\subsection{Origin of the Narrow Fe K$\alpha$ Line} \label{subsec:origin}

The origin of the narrow Fe K$\alpha$ line has long been in debate, with the typical location thought to be in either the outer accretion disc, BLR, or torus. Nearby AGN show a wide variety of measured line widths and no connection with the optical BLR as measured by the width of the H$\beta$ line, which suggests no universal location for the narrow Fe K$\alpha$ line \citep{Yaqoob2004,Nandra2006,Shu2010}. In our sample of seven nearby AGN, we use a new technique to probe the spatial extent of the emitting material, which provides an alternative measurement of the location of the narrow Fe K$\alpha$ line. We find that distance estimates from measurements of the intrinsic velocity width suggest BLR or torus origin, but that these measurements consistently underestimate the inferred spatial extent of the emitting material.

Many of the sources in our sample have previously shown evidence for extended Fe K$\alpha$ emission. Circinus, the closest object in our sample, shows Fe K$\alpha$ emission \citep{Marinucci2013, Arevalo2014} extending up to $\sim 3"$ ($\sim 60$ pc) from the nucleus, which is anti-correlated with molecular gas in the circumnuclear environment \citep{Kawamuro2019}. In NGC 1068, \citet{Young2001} revealed spatially extended Fe K$\alpha$ emission out to approximately 2.2 kpc from the nucleus, and \citet{Bauer2015} found that approximately 30\% of the Fe K$\alpha$ emission is extended on scales larger than $\sim 140$ pc. These two sources were also recently found to have the most extended emission in a sample of many bright AGN, using a combination of variability and imaging analyses \citep{Andonie2022}. Likewise, \citet{Guainazzi2012} found that the Fe K$\alpha$ emitting material extended up to 300 pc from the nucleus in Mrk 3, another highly obscured AGN in our sample. 

Extended emission is easier to detect in Compton-thick AGN because of the heavy obscuration of the nuclear emission, but it has also been detected in the Compton-thin AGN NGC 4388. Faint extended Fe K$\alpha$ emission was first detected out as far as 2.5 kpc in one direction by \citet{Iwasawa2003}, and more recently, additional observations have allowed for more detailed morphological studies of the extended Fe K$\alpha$ emission on scales of $\lesssim 1$ kpc \citep{Yi2021}. In NGC 4151, another Compton-thin AGN, the amount of extended Fe K$\alpha$ emission has been debated, with first evidence coming early on in the Chandra mission \citep{Ogle2000}. However, later \citet{Wang2011a} argued with updated MARX PSF calculations that the extended Fe K$\alpha$ emission could only account for $\sim 5\%$ of the total line emission. Likewise, \citet{Miller2018} and \citet{Szanecki2021} suggest that the narrow Fe K$\alpha$ line is produced at several hundred gravitational radii through modeling of first order HETG and Suzaku/NuSTAR spectra, respectively. Time delays also suggest a BLR origin, with \citet{Zoghbi2019} finding time delays between the Fe K$\alpha$ line and the continuum in the XMM-Newton data on the order of $\sim 3$ days.

We note also that a recent study focused on the developing a consistent picture of the torus of Circinus has implemented a similar analysis involving the first, second, and third order Chandra HETG data \citep{Uematsu2021}. Their analysis revealed a spatial extent of $\sigma_\theta = 0.66_{-0.21}^{+0.33}$ arcseconds, which is slightly larger than our results for Circinus of $\sigma_\theta = 0.37_{-0.10}^{+0.07}$ arcseconds. We suspect that our value is lower because \citet{Uematsu2021} considered both point-like emission from closer to the black hole and extended emission. They found that roughly 20\% of emission from the Fe K$\alpha$ line is extended in Circinus. Thus, our smaller extent is likely because our methodology can be interpreted as providing a spatially-averaged measurement of the extent of the Fe K$\alpha$ emitting material.

Our analysis provides another measure of spatial extent, avoiding the issues of pile-up and PSF contributions that must be corrected for and properly modeled in imaging analysis. Given that these issues greatly affect the nucleus in imaging data, our method is more sensitive to smaller extension than can be properly observed with imaging data. We find that the bulk of the Fe K$\alpha$ emitting material is extended on scales of $\sim 5-100$ pc across the seven sources in our sample, although with relatively large uncertainty due to the quality of the higher order HETG spectra. However, in most of our objects there is a discrepancy between the spatial extent of the material measured by $\sigma_\theta$ and the location of the gas measured with the intrinsic velocity width, $\sigma_v$, assuming motion is governed by the gravitational potential of the central SMBH. The intrinsic velocity of the gas indicates that the material is much closer to the black hole than we measure with the spatial extent. Here we present two possible explanations for the discrepancy between these distances.

Through modeling emission from both a point-like source and extended emission, \citet{Uematsu2021} found that extended emission in Circinus accounted for roughly 20\% of the Fe K$\alpha$ line emission. Our fitting method does not include these two features as separate model components, and instead can be interpreted as a spatial average of the two in a single model component. It is thus possible that our fits are being driven by gas closer to the BLR for the velocity width and by gas in and beyond the torus for the spatial extent. This suggests that the Fe K$\alpha$ emitting material responsible for the narrow Fe K$\alpha$ line in nearby AGN is present at a range of radii, from the inner BLR out to in and beyond the putative torus. This interpretation is consistent with recent findings that the variability and widths of the Fe K$\alpha$ line in a sample of almost 40 nearby AGN do not show a universal picture for the location of the narrow Fe K$\alpha$ line \citep{Andonie2022} and can help explain the discrepancy between different measurements of spatial extent.

Another possibility is that these large velocity widths are the result of non-gravitational motions driving the intrinsic velocity width to larger values. We investigate the nature of possible non-gravitational motions by allowing the redshift of the Fe K$\alpha$ line to vary, finding that six of the seven sources prefer slightly redshifted emission at roughly the 1$\sigma$ level. Assuming the Fe K$\alpha$ emitting material is Compton-thick so that our view is dominated by clouds on the far side of the nucleus, excess redshift would imply outflowing material. We note that this excess redshift could be the result of contamination from the Compton shoulder, which we have not included in our modeling. However, the intrinsic velocity widths we measure are much larger than the freefall velocities expected of inflowing gas in the galactic potential. Outflows can be driven to higher velocities via radiative, thermal, or magnetic launching mechanisms, and have been observed in UV and X-ray spectra with velocities ranging from $v \sim 100$s km s$^{-1}$ \citep[warm absorbers and their UV counterparts; e.g.][]{Reynolds1997,Crenshaw1999} up to $\gtrsim 0.1c$ \citep[ultra-fast outflows; e.g.][]{Tombesi2010,Tombesi2011,Tombesi2013,Gofford2013}. Outflows, with a variety of phases, scales, and velocities, have been detected in all seven objects in our sample (e.g. NGC 1068--\citealt{Cecil1990,Garcia-Burillo2014}; NGC 3783--\citealt{Kaspi2001,Mehdipour2017}; NGC 4151--\citealt{Crenshaw2007,Storchi-Bergmann2010}; NGC 4388--\citealt{Rodriguez-Ardila2017,Dominguez-Fernandez2020}; NGC 4507--\citealt{Fischer2013}; Mrk 3--\citealt{Ruiz2001,Gnilka2020}; Circinus--\citealt{Greenhill2003,Zschaechner2016,Stalevski2019}). Thus, we suggest that if the distance discrepancy is resolved by non-gravitational motions driving the intrinsic velocity widths higher, then it is likely that the Fe K$\alpha$ emitting material is outflowing from the central engine.

\subsection{Compton-Thick Sources} \label{subsec:comptonthick}

Two of the sources in our sample, NGC 1068 and Circinus, are well-known Compton-thick AGN (with $N_H \geq 1.5 \times 10^{24}$~cm$^{-2}$), and Mrk 3 is close to the Compton-thick limit. Extracting the intrinsic X-ray properties from Compton-thick AGN is difficult because the amount of emission transmitted directly from the AGN is often quite low compared to the amount of reprocessed emission. The additional scattering component that must be included in the modeling of Compton-thick sources is strongly dependent on the geometry of the system, which introduces a large amount of uncertainty. The best treatment of Compton-thick AGN to date involves analyzing data over a wide energy range. In particular, including hard X-ray data from observatories such as NuSTAR and Swift BAT is important to accurately model Compton-thick AGN because almost all of the emission with energy $\lesssim 10$~keV is scattered and reprocessed, leaving no intrinsic emission to detect in the soft X-ray band. Recent studies of NGC 1068 and Circinus have revealed estimates of the intrinsic AGN properties by fitting detailed models for various different geometries of the obscuring material to broad-band X-ray data from many different observatories \citep{Arevalo2014, Bauer2015,Marinucci2016,Uematsu2021}. 

Our simple absorption modeling of these sources fails to account for the additional scattering component and introduces some biases in the continuum parameter estimation. For NGC 1068 and Circinus, the column densities we find from our simple absorption model significantly underestimate those from the literature with a full model including both absorption and electron scattering (see Table \ref{tab:fitpars}). Thus, for these sources, we cannot properly estimate parameters like the equivalent width of the Fe K$\alpha$ line using our simple continuum modeling of the relatively narrow 4-9 keV band.

Properly modeling the absorption and scattering in these Compton-thick sources is beyond the scope of this work. Hence, we quote a range of values for the properties which depend on the intrinsic continuum, including the equivalent width and the subsequent gas mass and mass outflow rate measurements, for the two truly Compton-thick AGN in our sample (NGC 1068 and Circinus). The range is determined based on two extreme continuum estimates and utilizing the nuclear emission estimates from past multi-wavelength modeling, details of which are given in Section \ref{subsec:mass}. For Mrk 3, we find an average column density of $N_H \sim 0.8 \times 10^{24}$~cm$^{-2}$, which is at the low end of the line of sight values typically found in the literature \citep[e.g.][]{Yaqoob2015,Guainazzi2016}, but is in line with the finding that Mrk 3 is close to, but does not exceed the Compton-thick limit along our line of sight. Thus, we report values for Mrk 3, albeit with relatively high uncertainties, but urge caution in the interpretation of values that depend on the intrinsic continuum, such as the equivalent width, gas mass, and mass outflow rates, which may be overestimated due to our simple phenomenological modeling.

\subsection{Comparison to Molecular Gas} \label{subsec:compareALMA}

The Fe K$\alpha$ emission line traces cold, neutral gas around AGN, regardless of whether the gas is in atomic or molecular form. Molecular gas around AGN can also be traced via specific line transitions at submillimeter and infrared wavelengths, which have the advantage of high spatial resolution and velocity map information from spatially resolved spectroscopy. Studies of the molecular gas around nearby AGN have probed the circumnuclear region around the black hole in great detail, revealing significant outflows, indications of inflowing material \citep[see][and references therein]{Storchi-Bergmann2019}, and complexities of the molecular torus \citep[e.g.][]{Garcia-Burillo2016,Imanishi2018,Combes2019}. Molecular outflows can be quite massive, with mass outflow rates on the order of 100s-1000s $M_\odot$ yr$^{-1}$ \citep[e.g.][]{Feruglio2010,Sturm2011,Veilleux2013,Cicone2014,Tombesi2015}, which are similar to our derived values from the Fe K$\alpha$ emitting material. 

Two of the objects in our sample, NGC 1068 and Circinus, have been extensively studied with ALMA, allowing us to directly compare our results to the molecular gas properties. ALMA observations of NGC 1068 reveal significant molecular gas outflows on a variety of scales and gas densities with various molecular line tracers \citep[e.g.][]{Garcia-Burillo2014,Gallimore2016,Impellizzeri2019,Garcia-Burillo2019,Imanishi2020}. \citet{Garcia-Burillo2014} found significant molecular outflows in the circumnuclear disk (CND). On these scales of approximately 100 pc, the projected outflow velocities were on the order of $v \sim 100$ km s$^{-1}$ and the molecular outflow rate was $\dot{M}_\mathrm{out} \sim 63 M_\odot$ yr$^{-1}$. This is consistent with, but falls on the low end of, our large range of mass outflow rates for the neutral Fe K$\alpha$ emitting material in NGC 1068 (see Table \ref{tab:mass}). On smaller scales, similar to those traced by the neutral iron line in this study ($\sim 20$ pc), \citet{Garcia-Burillo2019} showed that about half of the mass in the molecular torus outflowing, although this portion of the outflow is significantly smaller in mass, momentum, and energy compared to the outflow in the CND. On even smaller scales of roughly a few parsecs, fast molecular outflows have been detected in both \citet{Gallimore2016} and \citet{Impellizzeri2019}, with projected velocities on the order of $v \sim 400-450$ km s$^{-1}$. Although with higher uncertainties, our results for NGC 1068 reveal similar mass outflow rates and higher velocities of the potentially outflowing Fe K$\alpha$ emitting material compared to those traced by molecular line transitions with ALMA.

Previous ALMA observations of the Circinus galaxy have revealed a molecular and atomic outflows, although the geometry of Circinus makes it difficult to distinguish between outflow and disk motions. \citet{Zschaechner2016} found that the molecular outflow in the central kiloparsec of Circinus had a consistent velocity with the ionized outflow \citep[$v \sim 150$ km s$^{-1}$][]{Veilleux1997}. However, this molecular outflow has significantly lower energetics than some of the massive molecular outflows \citep[e.g.][]{Cicone2014}, with an estimated mass outflow rate of $\dot{M}_\mathrm{out} \sim 0.35-12.3 M_\odot$ \citep{Zschaechner2016}. Like in NGC 1068, this is on the low end of our reported range of mass outflow rates for the Fe K$\alpha$ emitting material. On smaller scales, \citet{Izumi2018} found streaming motions indicative of an inflow to the AGN and an atomic outflow with a velocity on the order of $v \sim 60$ km s$^{-1}$. This outflow is thought to give rise to the thickness of the diffuse atomic gas and provides evidence for a multiphase, dynamic torus in Circinus. Additionally, \citet{Kawamuro2019} found that the spatial extent of the molecular line emission in Circinus was anti-correlated with the Fe K$\alpha$ emitting material, but had similar extents in the central 100 pc of the galaxy.

\subsection{Possibilities with Future X-ray Missions} \label{subsec:future}

This study opens the door for further work that can be done in tandem with future X-ray missions, such as XRISM, Athena, and Lynx. These missions will have microcalorimeters to perform non-dispersive, high resolution X-ray spectroscopy. The power of microcalorimeters in X-ray spectroscopy is evident from the Hitomi observations of NGC 1275 in the Perseus cluster, which provided the ability to resolve the narrow Fe K$\alpha$ line in NGC 1275 for the first time \citep{Hitomi2018}. In relation to this work, microcalorimeter spectroscopy will provide an independent measure of the intrinsic velocity width of the narrow Fe K$\alpha$ line and avoid contributions from the spatial extent, an effect that we have just shown has increased the line widths in dispersed Chandra HETG spectra. In principle, the spatial extent of the emitting material can be extracted by comparing Chandra HETG first order line widths with those from a microcalorimeter spectrum. This will be more prevalent in the future with X-ray missions, such as XRISM, Athena, and Lynx, but \citet{Reynolds2021b} have already demonstrated the possibilities of this technique by estimating the extent of the Fe K$\alpha$ emitting material in NGC 1275 with Hitomi SXS data and first order Chandra HETG data.

The higher order HETG spectra have significantly lower signal-to-noise compared to the first order spectrum due to their smaller effective areas, and thus are the limiting factor in our analysis and the cause of the large uncertainty on the spatial extent reported in this work. Therefore, being able to disentangle the spatial extent and intrinsic velocity width solely with the first order HETG spectrum and one from an X-ray microcalorimeter should greatly decrease the uncertainty in our current estimates of the spatial extent of the Fe K$\alpha$ emitting material in these objects, most of which are close to upper limits. Likewise, requiring adequate signal-to-noise in the higher order HETG spectra restricts the sample of AGN for which this analysis can be done to bright, nearby sources with an extensive amount of observation. As already shown in \citet{Reynolds2021b}, it is possible to obtain constraints on the Fe K$\alpha$ emitting material in NGC 1275, a difficult target because of the surrounding X-ray emitting intracluster medium of the Perseus cluster, by applying this technique to microcalorimeter and dispersive spectroscopy together. Thus, together with archival first order Chandra HETG data, future X-ray microcalorimeter spectroscopy will will also allow for this study to be extended to many more AGN.


\section{Conclusions} \label{sec:conclusion}

We have presented fits to the narrow Fe K$\alpha$ line in seven nearby AGN that include the contribution of spatial extent on the line width in HETG spectra. These AGN were chosen based on the criteria of having usable second and third order HETG spectra, which have effective areas which are approximately a factor of 15 lower than the first order data. Thus, in order to have sufficient data, these sources are bright, relatively local, and have been observed extensively with the HETG. Given their proximity, these sources are spatially extended on the Chandra ACIS-S detectors. The spatial extent has a larger contribution to the first order HETG data, making the first order line widths larger than the second and third order widths, as originally seen in \citet{Liu2016}. By fitting to all three orders simultaneously, we were able to break the degeneracy between the velocity width and spatial extent and fit directly to these values.

We find generally that the X-ray irradiated gas is extended on scales of approximately $~0.3-0.6$", corresponding to physical distances of the order $5-100$ pc. This indicates that the narrow Fe K$\alpha$ line is the result of reflection off of distant material. However, this conflicts with our measurements of distance using the velocity width, which gives distances on the order of $0.01-1$ pc. We suggest two possibilities to remedy this discrepancy. First, we propose that the difference could be the result of radially stratified emission with the emission closest to the black hole dominating the velocity width and the emission furthest from the black hole dominating the spatial extent. This scenario would imply that there is no universal location for the narrow Fe K$\alpha$ emitting material around AGN, but rather that the emission region extends from the broad line region out beyond the torus. Another possibility is that the discrepancy is due to non-gravitational motions driving the velocity width to larger values, namely in the form of outflows from the central engine. The argument for outflows is supported by large velocities relative to freefall in the galactic potential and slight excess redshift in the Fe K$\alpha$ line in the majority of our sources.

Using our measurement of the spatial extent of the emitting material, we estimate the mass of the emitting gas to be in the range of $10^5 - 10^8 M_\odot$. This gives potential mass outflow rates on the order 10-3000 $M_\odot$ yr$^{-1}$, assuming the velocity width of the Fe K$\alpha$ line is characteristic of the outflow velocity. These outflow rates, although highly uncertain, are comparable to molecular mass outflow rates as traced by specific line transitions observed with submillimeter observatories. We also investigate trends in our derived AGN properties, but find no significant trends given the large uncertainties on our results. In order to accurately determine if any trends exist between the extent of the Fe K$\alpha$ emission and various AGN properties, a larger sample of AGN is necessary. With future X-ray observatories, such as XRISM, coming online soon and providing non-dispersive, high resolution X-ray spectra and independent measures of the intrinsic velocity width of the narrow Fe K$\alpha$ line, this study could be extended to many more AGN with first order Chandra HETG spectra. 


\begin{acknowledgments}
We would like to thank the anonymous referee for helpful comments that improved the manuscript. We would also like to thank Dom Walton and Christos Panagiotou for their insightful feedback and discussion. M.M. is grateful for support from the Gates Cambridge Trust. C.S.R. thanks the STFC for support under the Consolidated Grant ST/S000623/1, as well as the European Research Council (ERC) for support under the European Union’s Horizon 2020 research and innovation programme (grant 834203). 
\end{acknowledgments}

\vspace{5mm}
\facility{CXO (HETG)}

\software{CIAO \citep{Fruscione2006},
XSPEC \citep{Arnaud1996}, 
MARX \citep{Davis2012},
Astropy \citep{Astropy2013, Astropy2018},
Matplotlib \citep{Hunter2007},
NumPy \citep{vanderWalt2011}}


\bibliography{extended_FeK_bib}{}
\bibliographystyle{aasjournal}


\appendix

\section{Fit Parameters for Each ObsID} \label{sec:append_tab}

Table \ref{tab:fitpars} shows the fit parameters for each ObsID for each object. The fit parameters are the column density of the source absorption, the photon index, the power law normalization, and the Gaussian line normalization. These are all fixed across all three orders.

\startlongtable
\begin{deluxetable*}{c c c c c c}
	
	\tablecaption{Fit parameters for each object and ObsID} \label{tab:fitpars} 
	
    \tablehead{\colhead{Object} & \colhead{ObsID} & \colhead{$N_H$} & \colhead{$\Gamma$} & \colhead{$K_\mathrm{power}$}  & \colhead{$K_\mathrm{line}$} \\
    & & \colhead{($10^{22}$~cm$^{-2}$)} & & \colhead{($10^{-2}$ photons~keV$^{-1}$~cm$^{-2}$ s$^{-1}$)} & \colhead{($10^{-4}$ photons~cm$^{-2}$ s$^{-1}$)} }

    \startdata
	NGC 1068\tablenotemark{$\dagger$} & 332 & $63.4_{-16.7}^{+18.5}$ & $2.41_{-0.49}^{+0.39}$ & $1.16_{-0.73}^{+1.05}$ & $0.85_{-0.23}^{+0.31}$ \\
    & 9148 & $72.6_{-15.6}^{+15.4}$ & $2.29_{-0.43}^{+0.35}$ & $1.13_{-0.66}^{+0.98}$ & $0.61_{-0.16}^{+0.20}$ \\
    & 9149 & $58.8_{-13.8}^{+14.9}$ & $2.30_{-0.38}^{+0.39}$ & $0.87_{-0.47}^{+0.95}$ & $0.70_{-0.16}^{+0.19}$ \\
    & 9150 & $73.8_{-21.0}^{+25.0}$ & $2.33_{-0.54}^{+0.47}$ & $1.08_{-0.73}^{+1.25}$ & $0.52_{-0.19}^{+0.30}$ \\
    & 10815 & $85.5_{-35.6}^{+49.5}$ & $2.33_{-0.70}^{+0.67}$ & $1.24_{-0.89}^{+1.53}$ & $0.94_{-0.45}^{+0.84}$ \\
    & 10816 & $103.4_{-47.3}^{+73.7}$ & $2.07_{-0.77}^{+0.72}$ & $1.16_{-0.88}^{+1.55}$ & $0.81_{-0.52}^{+1.21}$ \\
    & 10817 & $73.5_{-21.0}^{+24.1}$ & $2.26_{-0.50}^{+0.42}$ & $1.13_{-0.69}^{+1.05}$ & $0.92_{-0.33}^{+0.46}$ \\
    & 10823 & $78.0_{-19.7}^{+23.6}$ & $2.24_{-0.50}^{+0.41}$ & $1.14_{-0.72}^{+1.09}$ & $0.78_{-0.28}^{+0.40}$ \\
    & 10829 & $64.2_{-20.8}^{+25.3}$ & $2.49_{-0.56}^{+0.49}$ & $1.10_{-0.73}^{+1.32}$ & $0.55_{-0.19}^{+0.30}$ \\
    & 10830 & $53.1_{-18.1}^{+21.5}$ & $2.57_{-0.52}^{+0.45}$ & $1.11_{-0.71}^{+1.22}$ & $0.64_{-0.20}^{+0.29}$ \\
    \hline
    NGC 3783 & 2090 & $3.1_{-2.2}^{+2.1}$ & $1.67_{-0.12}^{+0.11}$ & $1.46_{-0.30}^{+0.34}$ & $0.38_{-0.07}^{+0.07}$ \\
    & 2091 & $8.6_{-2.2}^{+2.1}$ & $1.97_{-0.12}^{+0.11}$ & $2.71_{-0.56}^{+0.64}$ & $0.40_{-0.07}^{+0.08}$ \\
    & 2092 & $3.7_{-2.2}^{+2.1}$ & $1.75_{-0.12}^{+0.11}$ & $1.68_{-0.35}^{+0.39}$ & $0.43_{-0.07}^{+0.07}$ \\
    & 2093 & $7.4_{-1.9}^{+1.9}$ & $2.03_{-0.10}^{+0.10}$ & $4.09_{-0.77}^{+0.88}$ & $0.47_{-0.08}^{+0.08}$ \\
    & 2094 & $8.1_{-2.0}^{+2.0}$ & $1.99_{-0.11}^{+0.10}$ & $3.36_{-0.67}^{+0.75}$ & $0.38_{-0.07}^{+0.08}$ \\
    & 14991 & $8.0_{-3.7}^{+3.5}$ & $1.89_{-0.17}^{+0.15}$ & $1.76_{-0.51}^{+0.61}$ & $0.42_{-0.11}^{+0.12}$ \\
    & 15626 & $7.7_{-3.1}^{+3.0}$ & $1.73_{-0.15}^{+0.13}$ & $1.24_{-0.32}^{+0.39}$ & $0.36_{-0.08}^{+0.09}$ \\
    & 18192 & $9.8_{-3.2}^{+3.1}$ & $2.07_{-0.15}^{+0.14}$ & $2.69_{-0.73}^{+0.84}$ & $0.39_{-0.10}^{+0.12}$ \\
    & 19694 & $15.3_{-3.6}^{+3.3}$ & $2.24_{-0.17}^{+0.15}$ & $3.28_{-0.95}^{+1.11}$ & $0.37_{-0.10}^{+0.11}$ \\
    \hline
    NGC 4151 & 335 & $22.9_{-3.6}^{+3.4}$ & $1.67_{-0.17}^{+0.16}$ & $2.89_{-0.86}^{+1.08}$ & $1.77_{-0.27}^{+0.29}$ \\
    & 3052 & $8.3_{-1.7}^{+1.6}$ & $1.61_{-0.09}^{+0.09}$ & $4.74_{-0.82}^{+0.91}$ & $1.32_{-0.15}^{+0.15}$ \\
    & 3480 & $9.1_{-2.0}^{+2.0}$ & $1.54_{-0.11}^{+0.10}$ & $4.37_{-0.87}^{+0.99}$ & $1.39_{-0.20}^{+0.21}$ \\
    & 7829 & $20.6_{-4.5}^{+4.3}$ & $1.68_{-0.21}^{+0.17}$ & $1.51_{-0.53}^{+0.62}$ & $1.00_{-0.19}^{+0.21}$ \\
    & 7830 & $14.6_{-2.5}^{+2.3}$ & $2.02_{-0.13}^{+0.12}$ & $11.51_{-2.64}^{+3.07}$ & $1.21_{-0.26}^{+0.29}$ \\
    & 16089 & $24.7_{-2.2}^{+2.2}$ & $1.93_{-0.11}^{+0.11}$ & $4.79_{-1.01}^{+1.15}$ & $1.51_{-0.14}^{+0.15}$ \\
    & 16090 & $24.4_{-3.3}^{+3.1}$ & $1.68_{-0.16}^{+0.14}$ & $2.84_{-0.79}^{+0.95}$ & $1.38_{-0.20}^{+0.22}$ \\
    \hline
    NGC 4388 & 9276 & $37.2_{-5.1}^{+4.8}$ & $1.38_{-0.25}^{+0.24}$ & $0.75_{-0.31}^{+0.48}$ & $0.67_{-0.10}^{+0.11}$ \\
    & 9277 & $38.5_{-6.0}^{+5.7}$ & $1.71_{-0.30}^{+0.28}$ & $1.62_{-0.75}^{+1.28}$ & $0.76_{-0.14}^{+0.16}$ \\
    \hline
    NGC 4507 & 2150 & $64.7_{-5.8}^{+5.8}$ & $1.70_{-0.27}^{+0.27}$ & $2.61_{-1.14}^{+1.99}$ & $0.75_{-0.14}^{+0.16}$ \\
    \hline
    Mrk 3 & 873 & $72.1_{-10.4}^{+10.8}$ & $2.03_{-0.35}^{+0.33}$ & $0.99_{-0.51}^{+0.89}$ & $0.85_{-0.17}^{+0.20}$ \\
    & 12874 & $76.0_{-11.6}^{+12.0}$ & $1.98_{-0.37}^{+0.31}$ & $1.09_{-0.59}^{+0.88}$ & $0.86_{-0.19}^{+0.23}$ \\
    & 12875 & $81.7_{-18.4}^{+19.7}$ & $1.94_{-0.45}^{+0.40}$ & $1.08_{-0.65}^{+1.09}$ & $0.81_{-0.30}^{+0.41}$ \\
    & 13254 & $83.1_{-19.1}^{+22.6}$ & $1.98_{-0.50}^{+0.39}$ & $1.14_{-0.72}^{+1.10}$ & $0.76_{-0.28}^{+0.42}$ \\
    & 13261 & $68.7_{-20.8}^{+23.9}$ & $2.22_{-0.57}^{+0.46}$ & $1.20_{-0.81}^{+1.34}$ & $0.58_{-0.27}^{+0.40}$ \\
    & 13263 & $82.0_{-24.9}^{+32.7}$ & $1.98_{-0.57}^{+0.51}$ & $1.08_{-0.73}^{+1.38}$ & $1.03_{-0.44}^{+0.69}$ \\
    & 13264 & $82.7_{-16.1}^{+17.9}$ & $1.89_{-0.43}^{+0.37}$ & $1.03_{-0.60}^{+1.02}$ & $0.59_{-0.24}^{+0.32}$ \\
    & 13406 & $97.0_{-25.0}^{+30.2}$ & $1.85_{-0.52}^{+0.44}$ & $1.14_{-0.74}^{+1.20}$ & $1.42_{-0.51}^{+0.76}$ \\
    & 14331 & $76.0_{-15.5}^{+16.5}$ & $1.86_{-0.41}^{+0.38}$ & $0.85_{-0.49}^{+0.94}$ & $0.83_{-0.23}^{+0.29}$ \\
    \hline
    Circinus\tablenotemark{$\dagger$} & 10223 & $51.4_{-5.3}^{+5.4}$ & $1.26_{-0.17}^{+0.17}$ & $0.41_{-0.13}^{+0.15}$ & $3.87_{-0.33}^{+0.36}$ \\
    & 10224 & $50.4_{-6.1}^{+6.4}$ & $1.35_{-0.20}^{+0.18}$ & $0.46_{-0.16}^{+0.18}$ & $3.96_{-0.38}^{+0.43}$ \\
    & 10225 & $58.1_{-6.8}^{+7.0}$ & $1.56_{-0.21}^{+0.20}$ & $0.72_{-0.26}^{+0.31}$ & $4.43_{-0.47}^{+0.51}$ \\
    & 10226 & $79.5_{-13.7}^{+15.3}$ & $2.27_{-0.33}^{+0.30}$ & $3.66_{-1.78}^{+2.35}$ & $4.83_{-0.96}^{+1.20}$ \\
    & 10832 & $46.8_{-13.3}^{+14.3}$ & $0.64_{-0.33}^{+0.31}$ & $0.11_{-0.05}^{+0.07}$ & $3.92_{-0.74}^{+0.88}$ \\
    & 10833 & $59.1_{-11.0}^{+11.6}$ & $1.55_{-0.29}^{+0.27}$ & $0.77_{-0.35}^{+0.44}$ & $4.52_{-0.73}^{+0.86}$ \\
    & 10842 & $73.8_{-11.4}^{+12.4}$ & $2.01_{-0.28}^{+0.27}$ & $1.94_{-0.83}^{+1.12}$ & $5.30_{-0.82}^{+0.99}$ \\
    & 10843 & $54.2_{-8.0}^{+8.5}$ & $1.19_{-0.23}^{+0.21}$ & $0.35_{-0.13}^{+0.16}$ & $3.88_{-0.47}^{+0.52}$ \\
    & 10844 & $56.4_{-11.2}^{+11.7}$ & $1.73_{-0.29}^{+0.28}$ & $0.90_{-0.41}^{+0.50}$ & $3.40_{-0.60}^{+0.69}$ \\
    & 10850 & $53.0_{-16.5}^{+18.4}$ & $1.00_{-0.36}^{+0.36}$ & $0.25_{-0.13}^{+0.18}$ & $2.98_{-0.75}^{+0.94}$ \\
    & 10872 & $75.9_{-16.3}^{+17.6}$ & $1.55_{-0.34}^{+0.34}$ & $0.98_{-0.48}^{+0.64}$ & $5.54_{-1.22}^{+1.51}$ \\
    & 10873 & $67.4_{-13.2}^{+15.2}$ & $2.30_{-0.35}^{+0.33}$ & $2.88_{-1.45}^{+1.87}$ & $4.35_{-0.86}^{+1.09}$ \\
    & 62877 & $63.3_{-7.5}^{+7.7}$ & $1.67_{-0.23}^{+0.21}$ & $0.98_{-0.36}^{+0.44}$ & $3.84_{-0.46}^{+0.51}$ \\
    & 4770 & $64.6_{-7.9}^{+8.2}$ & $1.79_{-0.23}^{+0.22}$ & $1.19_{-0.46}^{+0.56}$ & $4.05_{-0.49}^{+0.54}$ \\
    & 4771 & $70.4_{-8.1}^{+8.6}$ & $1.64_{-0.23}^{+0.22}$ & $0.98_{-0.37}^{+0.46}$ & $4.61_{-0.56}^{+0.65}$ \\
    \enddata
    
    \flushleft
    \tablecomments{Columns are: (1) Object name, (2) Chandra ObsID, (3) Column density of source absorption, (4) Power law photon index, (5) Power law normalization, (6) Gaussian line normalization}
    \tablenotetext{\dagger}{The column densities ($N_H$) reported for NGC 1068 and Circinus underestimate the true column density along our line of sight due to our simple phenomenological modeling of absorption in a narrow energy band. See Section \ref{subsec:comptonthick} for details on how this affects our estimates of various physical parameters.}
    
\end{deluxetable*}

\section{Details of Equivalent Width Estimate in Equation 7} \label{sec:append_eqw}

The equivalent width of an emission line is defined as 
\begin{equation}
    \mathrm{EW} = \frac{\int_{E_\mathrm{min}}^{E_\mathrm{max}} \left( F - F_0 \right)\, dE}{F_0},
\end{equation}
where $F$ is the flux of the total spectrum (line and continuum), $F_0$ is the flux of the continuum at the line energy, and $E_\mathrm{min}$ and $E_\mathrm{max}$ bound the emission line. For the neutral Fe K$\alpha$ line, $F_0$ is the flux at 6.4~keV. The total flux of the Fe K$\alpha$ photons is given by $F - F_0$.

We can calculate the rate of Fe K$\alpha$ photon production, and therefore the photon flux of the Fe K$\alpha$ line, by counting fluorescing neutral iron atoms around the AGN. Consider an AGN with incident flux, $F_\mathrm{inc}$, on neutral iron atoms in the circumnuclear material. The incident photon flux is $F_\mathrm{inc} / h\nu = F_\mathrm{inc} / E$. If these photons have energy greater than $E_\mathrm{thresh} \approx 7.1$~keV, then they can photoionize neutral iron, removing an inner K shell electron. Then the rate of photoionization of an iron atom is given by
\begin{equation} \label{eq:ionizationrate}
    R_\mathrm{Fe, \, ion} = \int_{E_\mathrm{thresh}}^{\infty} \frac{F_\mathrm{inc}}{E} \sigma(E)\, dE,
\end{equation}
where $\sigma(E)$ is the photoionization cross section. Not all photoionized iron atoms will fluoresce and produce Fe K$\alpha$ photons. The probability of fluorescence is given by the fluorescent yield, $Y$, which for iron is $Y_\mathrm{Fe} = 0.33$. Then the rate of Fe K$\alpha$ photon production around the AGN is given by 
\begin{equation}
    R_{\mathrm{Fe\, K} \alpha} = N_\mathrm{Fe} Y_\mathrm{Fe} R_\mathrm{Fe,\, ion},
\end{equation}
where $N_\mathrm{Fe}$ is the total number of iron atoms. Approximating the iron in a shell of radius $R$, the number of iron atoms is 
\begin{equation}
    N_\mathrm{Fe} = Z_\mathrm{abs,\, Fe} N_H 4 \pi R^2,
\end{equation}
where $Z_\mathrm{abs,\, Fe}$ is the abundance of iron relative to hydrogen and $N_H$ is the column density of hydrogen. Therefore the rate of Fe K$\alpha$ photon production can be written as
\begin{equation}
    R_{\mathrm{Fe\, K} \alpha} = 4 \pi R^2 Z_\mathrm{abs,\, Fe} Y_\mathrm{Fe} N_H \int_{E_\mathrm{thresh}}^{\infty} \frac{F_\mathrm{inc}}{E} \sigma(E)\, dE.
\end{equation}

The photon flux of Fe K$\alpha$ photons at a radius $R$ is simply the rate of Fe K$\alpha$ photon production divided by $4 \pi R^2$. This can be related to the energy flux of Fe K$\alpha$ photons by
\begin{equation}
    F_{\mathrm{Fe\, K} \alpha} = E_{\mathrm{Fe\, K} \alpha} \frac{R_{\mathrm{Fe\, K} \alpha}}{4 \pi R^2} = E_{\mathrm{Fe\, K} \alpha} Z_\mathrm{abs,\, Fe} Y_\mathrm{Fe} N_H \int_{E_\mathrm{thresh}}^{\infty} \frac{F_\mathrm{inc}}{E} \sigma(E)\, dE,
\end{equation}
which is used to determine the equivalent width of the Fe K$\alpha$ line. We then assume that the photon index of the incident radiation from the AGN is $\Gamma = 2$, which yields a spectral index of $\alpha = 1$. This implies that the incident flux can be written as $F_\mathrm{inc} = A E^{-\alpha} \propto E^{-1}$, where $A$ is a constant. Then the equivalent width of the Fe K$\alpha$ line can be written 
\begin{equation}
    EW_{\mathrm{Fe\, K}\alpha} = \frac{E_{\mathrm{Fe\, K} \alpha} Z_\mathrm{abs,\, Fe} Y_\mathrm{Fe} N_H \int_{E_\mathrm{thresh}}^{\infty} A \frac{\sigma(E)}{E^2} \, dE}{A E_{\mathrm{Fe\, K}\alpha}^{-1}} \nonumber = E_{\mathrm{Fe\, K} \alpha}^2 Z_\mathrm{abs,\, Fe} Y_\mathrm{Fe} N_H \int_{E_\mathrm{thresh}}^{\infty} \frac{\sigma(E)}{E^2} \, dE.
\end{equation} \label{eq:eqw_theory}
This is Equation (1) of \citet{Reynolds2000}, for which the authors calculate $EW_\mathrm{Fe, \, max} \approx 65$ eV for NGC 4258. This calculation was based on a column density of $N_H = 10^{23}$~cm$^{-2}$ and solar value of $Z_\mathrm{abs,\, Fe}$. Hence, Equation \ref{eq:eqw} \citep[Equation (1) from][]{Hitomi2018} is simply a scaled version of Equation \ref{eq:eqw_theory}, which also accounts for possible clumpy or patchy material with the inclusion of the covering fraction, $f_\mathrm{cov}$.

\end{document}